%% file: arxiv_v2.tex
\documentclass[a4paper,11pt]{article}
\pdfoutput=1 

\usepackage{jheppub} 
\makeatletter
\DeclareRobustCommand*{\bfseries}{%
  \not@math@alphabet\bfseries\mathbf
  \fontseries\bfdefault\selectfont
  \boldmath
}
\makeatother

\usepackage{calc}
\usepackage{rotating}
\usepackage[english]{babel}
\usepackage[utf8]{inputenc}
\usepackage{graphicx}
\usepackage{subfig}
\usepackage{float}
\usepackage{amsmath}
\usepackage{amssymb}
\usepackage{amsthm}
\usepackage{latexsym}
\usepackage{dcolumn}
\usepackage{hyperref}
\input macros.tex

\newcolumntype{P}[1]{>{\centering\arraybackslash}p{#1}}

\restylefloat{figure}

\title{Qubit entanglement from forward scattering}

\author{Kamila Kowalska}
\author{and Enrico Maria Sessolo}

\affiliation{National Centre for Nuclear Research,\\
Pasteura 7, 02-093 Warsaw, Poland}

\emailAdd{kamila.kowalska@ncbj.gov.pl}
\emailAdd{enrico.sessolo@ncbj.gov.pl}

\abstract{In the context of entanglement in relativistic $2\to 2$ scattering described by a perturbative $S$-matrix, we derive analytically the concurrence for a mixed final state of two qubits corresponding to a discrete quantum number of the scattered particles. The qubit density matrix is obtained by tracing the momentum degrees of freedom out of the full density matrix of the scattered system. Given an initial product state, the derived concurrence depends at the leading order on the real part of the inelastic forward amplitude and the initial state only. We also point out that the real part of the forward amplitude provides a subleading correction to the linearized entropy, reducing it by an amount that, for a computational-basis state, is equivalent to the relative entropy of coherence. We illustrate our findings with two examples of phenomenological interest: high-energy scattering of two scalar fields in the two-Higgs doublet model, and high-energy electron-positron annihilation.
 }

\begin{document}
\maketitle

\setcounter{footnote}{0}

\section{Introduction\label{sec:intro}}

There has been growing interest in understanding the properties of quantum entanglement generated in relativistic
scattering processes. From the experimental point of view, the key question is 
how to detect and test entanglement between scattered particles at high-energy colliders like the LHC. This issue was analyzed for top-pair production\cite{Afik:2020onf,Fabbrichesi:2021npl,Severi:2021cnj,Afik:2022kwm,Aoude:2022imd,Aguilar-Saavedra:2022uye,Fabbrichesi:2022ovb,Severi:2022qjy,Dong:2023xiw,Varma:2023gwh,Aguilar-Saavedra:2023hss,Han:2023fci,ATLAS:2023fsd,Maltoni:2024tul,Aguilar-Saavedra:2024hwd,Duch:2024pwm}, for Higgs bosons decays\cite{Barr:2021zcp,Altakach:2022ywa,Ma:2023yvd,Ehataht:2023zzt,Altomonte:2024upf}, and in systems of massive gauge bosons\cite{Barr:2021zcp,Barr:2022wyq,Aguilar-Saavedra:2022wam,Ashby-Pickering:2022umy,Fabbrichesi:2023cev,Aoude:2023hxv,Bi:2023uop,Bernal:2023ruk,DelGratta:2025xjp,Goncalves:2025mvl,Goncalves:2025xer} (see also Ref.\cite{Barr:2024djo} for a recent review). In the same spirit, the idea of using entanglement to enhance the sensitivity of New Physics searches at the LHC was also explored\cite{Aoude:2022imd,Aoude:2023hxv,Maltoni:2024tul,Sullivan:2024wzl,Grossi:2024jae,DelGratta:2025qyp}. 

On the theoretical front, one primary goal would be to identify connections between the entanglement of particles scattered at colliders and some fundamental properties of the underlying quantum field theory, for example, to test the emergence of discrete, global, or gauge symmetries of the Lagrangian after some specific conditions are imposed on the entanglement of the final state. It was first observed in Ref.\cite{Beane:2018oxh} that entanglement minimization in elastic nucleon-nucleon scattering below the pion-production threshold implies the emergence of well-known spin-flavor symmetries. This finding may hint at the interesting possibility that the observed symmetries of particle physics have their root in a 
principle of entanglement suppression\cite{Beane:2018oxh,Beane:2020wjl,Low:2021ufv,Beane:2021zvo,Liu:2022grf,Hu:2025lua}. 
However, while this idea seems to find a positive realization in nuclear theory and low-scale quantum chromodynamics~(QCD), in high-energy particle physics we are still far from definite conclusions. Several theories were investigated in this context (including quantum electrodynamics~(QED)\cite{Cervera-Lierta:2017tdt}, QCD\cite{Garrido:2025xpk}, the two-Higgs doublet model~(2HDM)\cite{Carena:2023vjc,Kowalska:2024kbs,Chang:2024wrx,Carena:2025wyh,Busoni:2025dns}, the quark sector of the Standard Model~(SM)\cite{Thaler:2024anb}), with no common conclusion regarding whether the symmetry-generating entanglement should be minimized or maximized, or whether any symmetry emerges at all from such an extremization principle. 

In an attempt to uncover possible connections between entanglement and symmetries in particle physics, it is important to specify what kind of entanglement one is really interested in. While any non-trivial interaction of two scattering particles results in entanglement of 
different modes of their momenta, in a way that is proportional to the total cross section\cite{Balasubramanian:2011wt,Hsu:2012gk,Seki:2014cgq,Peschanski:2016hgk,Carney:2016tcs,Faleiro:2016lsf,Peschanski:2019yah}, quantum correlations between discrete quantum numbers (spin, polarization, flavor, etc.) strongly depend on the underlying properties of the $S$-matrix. It is thus justified to expect that the properties of the Lagrangian, like, \textit{e.g.}, possible symmetries, would leave a mark on the entanglement of these \textit{qudit} degrees of freedom.
In order to quantify this type of entanglement, one must be able to sift the momentum degrees of freedom out of the final-state density matrix. 
In general, this is done in one of two alternative ways. 

In one approach, the final-state density matrix is {\it projected} onto a specific outgoing momentum vector, usually parameterized by a single scattering angle. In this case, the scattered bi-partite state is pure and standard entanglement measures can be used to quantify the entanglement of the qudits at any given element of solid angle. By construction, 
in this setup probability is not conserved between the initial and final state, a feature that may obscure the connection between the amount of entanglement generated in the scattering event and the properties of the underlying theory. 
On the other hand, this is exactly the kind of entanglement that can be (and actually is) measured experimentally.

An alternative approach to quantify qubit entanglement was proposed in our previous article\cite{Kowalska:2024kbs}. Instead of projecting the final-state density matrix onto the fixed outgoing momenta, the momentum degrees of freedom are traced out. Since in this framework the unitarity of the $S$-matrix is guaranteed at the required order of the perturbative expansion by the optical theorem, it is possible to quantify the amount of entanglement generated or lost in the interaction.  
Unlike the projection case, in this approach the momentum-reduced bi-partite scattered state is usually mixed. 
Hence, one must employ dedicated measures like the \textit{concurrence}\cite{Hill:1997pfa,Wootters:1997id} or the PPT criterion\cite{PhysRevLett.77.1413,HORODECKI19961}
to compute the entanglement of the final-state qubits, while measures like the von Neumann (or linearized) entropy quantify the entanglement (or better, the non-separability) between the momentum and qubit degrees of freedom. Interesting phenomena can emerge, for example a flow of entanglement between the different partitions of the Hilbert space\cite{Kowalska:2024kbs}.

In this study, we take a closer look at the origin of entanglement between two qubits in the final state of a relativistic scattering event described by a perturbative $S$-matrix. Building on the findings of Ref.\cite{Kowalska:2024kbs}, we derive an analytic formula for the concurrence of a momentum-reduced density matrix in terms of the scattering amplitude and the initial-state qubit. This leads to an interesting observation that, to our knowledge, has not been pointed out before: the entanglement of two qubits is generated at the leading order by the real part of the \textit{forward} scattering amplitude, in contrast to the entanglement between the momentum and qubit partitions, 
which at the leading order stems from the imaginary part or, equivalently -- through the optical theorem -- the total cross section. Moreover, even when considering the generation of momentum-qubit entanglement, we show that the real part of the forward amplitude provides an important subleading correction, whose size is constrained by the requirement of unitarity.  
In the context of a perturbative expansion of the $S$-matrix, our results hint at the possibility of categorizing, for different theories, the strength of post-scattering qubit entanglement on the basis of stereographic properties of the amplitude. 

The paper is organized as follows. In \refsec{sec:set} we revisit the formalism introduced in Ref.\cite{Kowalska:2024kbs} and recast it an a compact matrix form which can be applied to a generic qudit case. In \refsec{sec:entpow} we use this formalism to derive an expression for the linearized entropy of the scattered final state in terms of the relative entropy of coherence of the density matrix.  
In \refsec{sec:conc} we focus on entanglement between qubits and we present an analytic expression for the concurrence 
in terms of the scattering amplitude and the initial qubit state. We show that, in the case of an initial product state,
the derived expression acquires a simple form. Section~\ref{sec:examp} presents two practical applications of the derived concurrence formula: high-energy scattering of two flavored scalars, and high-energy $e^+ e^-$ annihilation in QED. We summarize our findings in \refsec{sec:sum}.
The appendices feature, respectively, a generalization of some of the formulas in Secs.~\ref{sec:set} and \ref{sec:entpow} to the case of wave packets, some intermediate results about the size of the eigenvalues of the density matrix and, finally, 
the detailed proof of the concurrence formula presented in \refsec{sec:conc}. 

\section{The setup\label{sec:set}}

We begin by recalling some basic notions of entanglement in perturbative $2\to 2$ scattering. 
We follow closely the formalism of our previous work\cite{Kowalska:2024kbs}, which we recast here in a more compact matrix form. Consider two scattering particles that carry, beside momentum degrees of freedom, a discrete-index quantum number. The latter can be identified, for example,  with helicity\cite{Cervera-Lierta:2017tdt,Fonseca:2021uhd,Blasone:2024dud,Blasone:2024jzv,Blasone:2025tor,Blasone:2025ddi}, polarization\cite{Garrido:2025xpk,Fedida:2022izl}, or even flavor\cite{Carena:2023vjc,Kowalska:2024kbs,Chang:2024wrx,Carena:2025wyh}. If the discrete-index dimension is equal~2 we can interpret it as a quantum information theoretical qubit. In general, a discrete quantum number with $d$ measured values can be considered, thus  extending the notion of a qubit into a broader qudit concept.

The total Hilbert space  of the scattering particles is constructed as 
\be
\mathcal{H}_{\textrm{tot}}=\mathcal{H}_{\textrm{mom}}\otimes\mathcal{H}_{\textrm{qd}}=L^2(\mathbb{R}^3\otimes \mathbb{R}^3)\otimes \mathbb{C}^{d^2}\,,
\ee
where $\mathcal{H}_{\textrm{mom}}=L^2(\mathbb{R}^3\otimes \mathbb{R}^3)= L^2(\mathbb{R}^3)\otimes L^2(\mathbb{R}^3)$ is the momentum Hilbert space of the two scattering particles, while the Hilbert space of the two qudits is given by $\mathcal{H}_{\textrm{qd}}=\mathbb{C}^{d^2}=\mathbb{C}^d\otimes \mathbb{C}^d$. We denote the  
basis for $\mathcal{H}_{\textrm{tot}}$ as $|\mathbf{p}_1 \mathbf{p}_2 \rangle |\alpha\beta\rangle$, where $\mathbf{p}_1, \mathbf{p}_2$ are continuous 3-dimensional variables and $\alpha,\beta$ can take values that range from 1 to $d$. 

The momentum-basis vectors are normalized as
\be\label{eq:norconv}
\langle\mathbf{p}_1\mathbf{p}_2 | \mathbf{p}_1' \mathbf{p}_2' \rangle
= (2\pi)^6\,4 E_{1} E_{2}\, 
\delta^3(\mathbf{p}_1-\mathbf{p}_1')\, \delta^3(\mathbf{p}_2-\mathbf{p}_2')\,,
\ee
where $E_1$, $E_2$ denote the $0$-th components of momentum 4-vectors $p_{1,2}=(E_{1,2},\mathbf{p}_{1,2})$ in Minkowski space. Given \refeq{eq:norconv} one can introduce
an indeterminate ``volume'' factor 
\be\label{eq:volume}
V=\langle\mathbf{p}_1\mathbf{p}_2 | \mathbf{p}_1 \mathbf{p}_2 \rangle = 4 E_1 E_2\left[(2\pi)^3\,\delta^3(0)\right]^2,
\ee
which is due to the fact that the basis vectors do not belong to the momentum Hilbert space 
$L^2(\mathbb{R}^3\otimes \mathbb{R}^3)$, but are instead tempered distributions acting on a dense subspace of $\mathcal{H}_{\textrm{mom}}$.
The computational-basis vector for a two-qudit state is normalized trivially as
\be\label{eq:normfl}
\langle\alpha\beta | \gamma\delta\rangle=\delta_{\alpha\gamma}\,\delta_{\beta\delta}\,.
\ee

In the following we shall limit ourselves to initial two-particle states before scattering that consist 
of plane waves with a highly collimated momentum distribution around initial values~$\mathbf{p}_A,\mathbf{p}_B$:
\be\label{eq:instate}
|\textrm{in}\rangle =\frac{1}{\sqrt{V}}\,\sum_{\alpha,\beta=1}^d a_{\alpha\beta}|\mathbf{p}_A\mathbf{p}_B\rangle |\alpha\beta\rangle\,,
\ee
where $|\mathbf{p}_A\mathbf{p}_B\rangle$ is an element of the basis and $\sum_{\alpha,\beta}|a_{\alpha\beta}|^2=1$.
The $|\textrm{in}\rangle$ state is separable between momentum and qudit but, importantly, the two qudits may be entangled.  

In the basis $\{\langle \mathbf{p}_i \mathbf{p}_j|\langle \gamma\delta |$, $|\mathbf{p}_a \mathbf{p}_b \rangle |\alpha\beta\rangle\}$ the elements of the $S$-matrix are defined as
\begin{multline}\label{eq:Smat}
S_{\gamma\delta\alpha\beta}^{ijab}=\left(\mathcal{I}+iT\right)_{\gamma\delta\alpha\beta}^{ijab}\\
=(2\pi)^6\,4\,E_a E_b\, \delta_{\gamma\delta\alpha\beta}^{ijab}
+(2\pi)^4\delta^4(p_a+p_b-p_i-p_j)\,
i\mathcal{M}_{\gamma\delta,\alpha\beta}(p_a,p_b\to p_i,p_j)\,,
\end{multline}
where 
\be
\delta_{\gamma\delta\alpha\beta}^{ijab}=\delta_{\alpha\gamma}\,\delta_{\beta\delta}\,\delta^3(\mathbf{p}_a-\mathbf{p}_i)\,\delta^3(\mathbf{p}_b-\mathbf{p}_j)\,,
\ee
and the scattering amplitude $\mathcal{M}$ carries discrete indices 
$\alpha\beta\gamma\delta$ besides momentum indices. Using Eqs.~(\ref{eq:instate}) and (\ref{eq:Smat}), the post-scattering final-state can be written as
\begin{multline}\label{eq:outst}
|\textrm{out}\rangle = S|\textrm{in}\rangle=\frac{1}{\sqrt{V}}\,\sum_{\alpha,\beta=1}^d a_{\alpha\beta}|\mathbf{p}_A\mathbf{p}_B\rangle |\alpha\beta\rangle\\
+\frac{i}{\sqrt{V}}\sum_{\alpha,\beta,\gamma,\delta}  a_{\alpha\beta}\,   |\gamma\delta\rangle  \int d\Pi_2\,\mathcal{M}_{\gamma\delta,\alpha\beta}(p_A,p_B\to p_i,p_j)\,  |\mathbf{p}_i \mathbf{p}_j\rangle\,,
\end{multline}
where we introduced the common short-hand notation for the phase-space integral,
\be\label{eq:delmax}
\int d\Pi_2\equiv  \int \int \frac{d^3 p_i}{(2\pi)^3}\frac{1}{2 E_i}\frac{d^3 p_j}{(2\pi)^3}\frac{1}{2 E_j}\,
(2\pi)^4\delta^4(p_A+p_B-p_i-p_j)\,.
\ee
Denoting 
\be
\aket=\sum_{\alpha,\beta}a_{\alpha\beta} |\alpha\beta\rangle
\ee
and dropping the explicit dependence of the amplitude on momentum 
and discrete indices, one can recast \refeq{eq:outst} in a compact matrix form,
\be\label{eq:finstat}
|\textrm{out}\rangle=\frac{1}{\sqrt{V}}\aket |\mathbf{p}_A\mathbf{p}_B\rangle+\frac{i}{\sqrt{V}}\int d\Pi_2\, \mathcal{M}\aket |\mathbf{p}_i \mathbf{p}_j\rangle\,.
\ee

The post-scattering density matrix is given by
\begin{multline}\label{eq:outrho}
|\textrm{out}\rangle\langle\textrm{out}|=\frac{1}{V}\pa\otimes|\mathbf{p}_A\mathbf{p}_B\rangle\langle\mathbf{p}_A\mathbf{p}_B|\\
+\frac{i}{V}\int d\Pi_2\left(\mathcal{M}\pa\otimes|\mathbf{p}_i\mathbf{p}_j\rangle\langle\mathbf{p}_A\mathbf{p}_B|-\pa\mathcal{M}^\dagger \otimes|\mathbf{p}_A\mathbf{p}_B\rangle\langle\mathbf{p}_i\mathbf{p}_j|\right)\\
+\frac{1}{V}\int d\Pi_2\int d\Pi_2\left(\mathcal{M}\pa\mathcal{M}^\dagger\right)\otimes |\mathbf{p}_i\mathbf{p}_j\rangle\langle\mathbf{p}_k\mathbf{p}_l|\,,
\end{multline}
where we have defined the projector onto the initial-state vector, $\pa\equiv \aket\abra$.

Using Eqs.~(\ref{eq:volume}) and~(\ref{eq:finstat}) it is easy to verify that the scattered state is correctly normalized via the optical theorem (cf.~Eq.~(2.12) of Ref.\cite{Kowalska:2024kbs}),
\be\label{eq:ot}
\langle\textrm{out}|\textrm{out}\rangle= 1+\Delta \abra\left(-2\, \textrm{Im}\,\mfw+\int d\Pi_2\, \mathcal{M}^{\dag}\mathcal{M} \right)\aket,
\ee 
where $\mathcal{M}_{\textrm{fw}}\equiv \mathcal{M}(p_A,p_B\to p_A, p_B)$ is the amplitude in the forward direction and 
\be
\Delta=\frac{(2\pi)^4\delta^4(0)}{4 E_A E_B\left[(2\pi)^3\,\delta^3(0)\right]^2}
\ee
is an indeterminate normalization factor of the momentum space introduced in Ref.\cite{Kowalska:2024kbs}.

\section{Linearized entropy\label{sec:entpow}}

In this study we seek to quantify entanglement between the discrete degrees of freedom after the entire momentum space has been traced out. The momentum-reduced density matrix, $\rho_{Q}=\textrm{Tr}_p( |\textrm{out}\rangle \langle\textrm{out}|)$, is obtained from \refeq{eq:outrho},
\be\label{rho:flav}
\rho_Q=\pa+i\Delta\left(\mfw\mathcal{P}_A-\pa\mfw^{\dag}\right)+\Delta\int d\Pi_2\left(\mathcal{M}\pa \mathcal{M}^\dag\right),
\ee
cf.~Eq.~(2.16) in Ref.\cite{Kowalska:2024kbs}.\footnote{In the projection approach that was mentioned in the Introduction, the qudit density matrix is obtained by projecting the final-state density matrix $|\textrm{out}\rangle\langle\textrm{out}|$ onto a fixed momentum vector, which we denote $\frac{1}{\sqrt{V}}|\mathbf{p}_C\mathbf{p}_D\rangle$. The projected state is then given by
\be
|\textrm{proj}\rangle=\frac{1}{V}|\mathbf{p}_C\mathbf{p}_D\rangle\langle\mathbf{p}_C\mathbf{p}_D|\textrm{out}\rangle\,,
\ee
from which the qudit density matrix follows,
\be\label{eq:rhopr}
\rho_{\textrm{proj}}=\frac{|\textrm{proj}\rangle\langle\textrm{proj}|}{\langle\textrm{proj}|\textrm{proj}\rangle}=\frac{\mathcal{M}(\mathbf{p}_C,\mathbf{p}_D)\pa\mathcal{M}(\mathbf{p}_C,\mathbf{p}_D)^\dagger}{\abra\mathcal{M}(\mathbf{p}_C,\mathbf{p}_D)^\dagger\mathcal{M}(\mathbf{p}_C,\mathbf{p}_D)\aket}\,.
\ee
We reiterate that qubit entanglement derived from \refeq{eq:rhopr} has a different interpretation and different numerical value than qubit entanglement that will be discussed in \refsec{sec:conc}.}

Equation~(\ref{rho:flav}) applies to pure and mixed states alike. To quantify the separability between the momentum and qudit Hilbert spaces, we calculate the \textit{linearized entropy},
\be\label{eq:def}
\mathcal{E}=1-\textrm{Tr}(\rho_Q^2)\,.
\ee
Expanding the trace at order $\mathcal{O}(\mathcal{M}^2)$ and
employing \refeq{eq:ot}, one obtains 
\begin{multline}\label{eq:linen}
\mathcal{E}=2\Delta \abra\int d\Pi_2\left(\mathcal{M}^{\dag}\mathcal{M}- \mathcal{M}\mathcal{P}_A\mathcal{M}^{\dag} \right)\aket\\
-\,\Delta^2\abra\left(2\mfw^{\dag}\mfw-\mfw\pa\mfw -\mfw^\dagger\pa\mfw^\dagger\right)\aket\,,
\end{multline}
cf.~Eq.~(2.18) in Ref.\cite{Kowalska:2024kbs}.

It is instructive to take a closer look at \refeq{eq:linen}. As was discussed in Ref.\cite{Kowalska:2024kbs}, 
the indeterminate normalization $\Delta$ is subject to an upper bound, $\Delta \leq \Delta_{\textrm{max}}\ll 1$, 
which enforces the positivity of the entropy (a derivation of the upper bound is presented in Appendix~\ref{app:WP}). This means that \refeq{eq:linen} can be treated perturbatively in $\Delta$ and that
the upper line yields the leading term in the expansion. 
By using the completeness of the Hilbert space and relating the amplitude to the cross section, $(\mathcal{M}^{\dag}\mathcal{M})\,d\Pi_2=4 E_a E_b |\mathbf{v}_A-\mathbf{v}_B|\, d\sigma$, one can recast \refeq{eq:linen} as\cite{Peschanski:2016hgk,Peschanski:2019yah,Low:2024mrk,Low:2024hvn,Sou:2025tyf} 
\be\label{eq:arealaw}
\mathcal{E}=4\,\Delta\, s\left(\sigma_{\aket\to\textrm{all}}-\sigma_{\aket\to\aket}\right)+\mathcal{O}(\Delta^2)\,,
\ee
where $\sigma_{\aket\to\textrm{all}}$ and $\sigma_{\aket\to\aket}$ are the total and ``elastic'' scattering cross sections of the $|\textrm{in}\rangle$ state and we have taken the high-energy limit with $s=4 E_A E_B$ being the usual Mandelstam variable in the c.o.m.~frame. This property, dubbed as ``the area law'' in the literature\cite{Low:2024mrk,Low:2024hvn}, states 
that, at the linear order in $\Delta$, the linearized entropy is proportional to the inelastic cross section for the initial qudit state \aket\ going to anything but itself. Similarly, entanglement between two scattered particles (thus including momentum degrees of freedom) is proportional to the total scattering cross section.

On the other hand, we can point out some interesting properties of the second line in \refeq{eq:linen}, despite it being suppressed by one perturbation order in $\Delta$. First we note that, since $\rho_Q$, $\pa$ are Hermitian and $\Delta$ is real, 
all three pieces in    
\refeq{rho:flav} are Hermitian and their diagonal is real. 
It follows that, while the imaginary part of the forward amplitude ($\textrm{Im}\,\mfw$, which appears at the 1-loop level first) can belong to the diagonal part of $i\Delta\,(\mfw\mathcal{P}_A-\pa\mfw^{\dag})$, the real part ($\textrm{Re}\,\mfw$, appearing at the tree level)
belongs to the off-diagonal part of $\rho_Q$. It is thus a \textit{coherence} of the density matrix, \textit{i.e.},  
a measure of quantum superposition in a given basis. When $|A\rangle$ belongs to the computational basis it is easy to prove\footnote{Let us consider a state in the computational basis, $|A\rangle=|\hat{\alpha}\hat{\beta}\rangle$, and define
$d^2$ projectors $P_{\alpha\beta}=|\alpha\beta\rangle \langle \alpha \beta|$. 
Decompose $\rho_Q$ into a diagonal and an off-diagonal part, $\rho_Q=\rho_{\textrm{diag}}+\rho_{\textrm{off}}$, 
and express the diagonal part in terms of the $P_{\alpha\beta}$ projectors,
\be\label{eq:rhod}
\rho_{\textrm{diag}}=\sum_{\alpha,\beta} P_{\alpha\beta}\,\rho_Q\, P_{\alpha\beta}=\pa-2 \Delta\, \textrm{Im}\left(\pa \mfw \pa \right)  +\sum_{\alpha,\beta}\Delta \int d\Pi_2 P_{\alpha\beta} \left(\mathcal{M}\pa\mathcal{M}^{\dag}\right) P_{\alpha\beta}\,.
\ee
After squaring \refeq{eq:rhod} and expanding to order $\mathcal{O}(\mathcal{M}^2)$ one gets
\be
\rho_{\textrm{diag}}^2=\pa-4\Delta \, \textrm{Im}\left(\pa \mfw \pa \right) +2\Delta \int d\Pi_2 \pa\left(\mathcal{M}\pa\mathcal{M}^{\dag}\right)\pa+\mathcal{O}(\mathcal{M}^4)\,,
\ee
which, after applying the optical theorem, leads to 
\be
\textrm{Tr}(\rho_{\textrm{diag}}^2)=1-2\Delta \abra\int d\Pi_2\left(\mathcal{M}^{\dag}\mathcal{M}- \mathcal{M}\mathcal{P}_A\mathcal{M}^{\dag} \right)\aket\,.
\ee
} 
that the $\mathcal{O}(\Delta^2)$ term in \refeq{eq:linen} is none other than, in the language of quantum information theory, the \textit{relative entropy of coherence}, defined as\cite{Baumgratz:2013ecx}
\bea
S_{\textrm{rel}}(\rho_Q)&\equiv &\textrm{Tr}(\rho^2_Q)-\textrm{Tr}(\rho_{\textrm{diag}}^2)\nonumber\\
 &=&\Delta^2\abra\left(2\mfw^{\dag}\mfw-\mfw\pa\mfw -\mfw^\dagger\pa\mfw^\dagger\right)\aket\,,
\eea
where $\rho_{\textrm{diag}}$ is the diagonal part of $\rho_Q$ (in a given basis). 

At each order in the perturbative expansion, the real part of the forward amplitude contributes to the 
$\Delta^2$ piece of \refeq{eq:linen}, reducing the amount of entanglement between momenta and qudits. 
The actual size of this reduction depends on the parameter $\Delta$. As we show in Appendix~\ref{app:WP}, where we work with generic $L^2(\mathbb{R}^3)$ wave packets of initial-state momenta, $\Delta$ stands as a proxy for the overlap in solid angle of the momentum distribution of the initial and final states. As was mentioned above, for initial wave packets peaked sharply around the incoming momenta, $\Delta$ is perturbatively small and the $\Delta^2$ contribution in \refeq{eq:linen} can be neglected.   

On the other hand, this is not the case for wave packets with a finite solid-angle spread 
centered about $\mathbf{p}_{A,B}$. 
In particular, for complete overlap of the initial and final state (\textit{e.g.}, the $s$-wave states discussed in Ref.\cite{Chang:2024wrx}) $\Delta$ reaches its maximal value. This coincides with an upper bound resulting from unitarity of the density matrix, $\textrm{Tr}(\rho_Q^2)\leq 1$.

As we show in \refeq{eq:max_over} of Appendix~\ref{app:WP}, the actual value of $\Delta_{\textrm{max}}$ depends on the 
angular distribution of the scattering amplitude and may differ theory by theory. For example, in the case of scalar scattering analyzed in Ref.\cite{Kowalska:2024kbs}, where $\mathcal{M}$ is isotropic, $\Delta_{\textrm{max}}=1/(16\pi)$; in the case of $e^+ e^-$~annihilation, discussed in \refsec{sec:QED}, it is $\Delta_{\textrm{max}}=1/(6\pi)$.
Interestingly, for $\Delta=\Delta_{\textrm{max}}$ the linearized entropy~(\ref{eq:linen}) vanishes and the momentum and qudit Hilbert spaces remain separable even after the scattering event takes place. This also means that the momentum-reduced density matrix~(\ref{rho:flav}) describes in such case a pure quantum state.

\section{Concurrence for a mixed state\label{sec:conc}}

The momentum-reduced density matrix of \refeq{rho:flav} describes a quantum state which is, in general, mixed. While the linearized entropy (\ref{eq:def}) can be readily applied to a mixed two-qudit system, capturing separability of the momentum and qudit Hilbert spaces, it is less trivial to quantify the entanglement between the two qudits. 

In the physically important case of two qubits~($d=2$), an
appropriate measure of entanglement is the concurrence\cite{Hill:1997pfa,Wootters:1997id}. It is defined as
\be\label{eq:conc}
C(\rho)=\max\{0,\lam_1-\lam_2-\lam_3-\lam_4\}\,,
\ee
where $\lam_i$ are the square roots of the non-negative eigenvalues (ordered from largest to smallest) of the Hermitian matrix $R=\rho\tilde{\rho}$, where $\tilde{\rho}$ is the spin-flipped state of $\rho$, defined as
\be
\tilde{\rho}=\Sigma\,\rho^\ast\Sigma\,,
\ee
where we define $\Sigma=\sigma_y\otimes\sigma_y$, and $\sigma_y$ indicates the second Pauli matrix. 

Unlike the linearized entropy in \refeq{eq:linen},
in the form given in \refeq{eq:conc} the concurrence does not yield much physical insight on possible relations between the entanglement generated in a scattering event and the properties of the $S$-matrix. Therefore, in this section we are going to recast $C(\rho)$ in a form that will make those relations clearer: in terms of the initial qubit state and the scattering amplitude. 

We first write the spin-flipped matrix $\tilde{\rho}_{Q}$ in a form analogous to $\rho_Q$ in \refeq{rho:flav}:
\be\label{eq:sigr}
\tilde{\rho}_{Q}=\prb-i\Delta\left(\mtilfw\prb-\prb\,\mtilfw^\dagger\right)+\Delta\int d\Pi_2\,\mtil\prb\mtil^\dagger\,,
\ee
where 
\be
\mtil=\Sigma\mathcal{M}^\ast\Sigma,\qquad \mtilfw=\Sigma\mfw^\ast\Sigma
\ee
denote the spin-flipped scattering amplitudes. The spin-flipped initial state is given by 
\be
 \bket=\Sigma|A^{\ast}\rangle\,, \qquad \prb=\bket\bbra,\qquad \langle B|B\rangle=1\,.
\ee

Let us next introduce the complex quantity
\be\label{eq:cadef} 
c_A \equiv \bbra A\rangle\,,
\ee
in analogy with the form of the concurrence of any \textit{pure} state $|P\rangle$\cite{Hill:1997pfa,Wootters:1997id},
\be
C(|P\rangle)=|\langle \tilde{P}|P\rangle|\,,
\ee
where $|\tilde{P}\rangle$ is the spin-flipped state of $|P\rangle$.

The matrix $R_Q=\rho_Q\tilde{\rho}_{Q}$ can be now straightforwardly obtained from \refeq{rho:flav} and \refeq{eq:sigr}. The actual formula is lengthy, yet not very illuminating, and we do not show it explicitly. Importantly, though, the matrix $R_Q$ has at most two dominant non-zero eigenvalues (see Appendix~\ref{app:2eig} for a proof), let us denote them $\xi_1$ and $\xi_2$. One can thus write 
\be\
C(\rho_Q)=\left|\sqrt{\xi_1}-\sqrt{\xi_2}\right|\,.
\ee
Using $\textrm{Tr}(R_Q)=\xi_1+\xi_2$ and $\textrm{Tr}(R_Q^2)=\xi_1^2+\xi_2^2$, the square of the concurrence can be expressed as
\be\label{eq:contr}
C^2(\rho_Q)=\textrm{Tr}(R_Q)-\sqrt{2\left[\textrm{Tr}(R_Q)^2-\textrm{Tr}(R_Q^2)\right]}\,.
\ee
In Appendix~\ref{app:con}, we use \refeq{eq:contr} as a starting point to derive a formula for the concurrence as a function of the scattering amplitude, valid in a perturbative expansion up to the order $\mathcal{O}(\mathcal{M}^2)$. 
The final expression reads 
\begin{multline}\label{eq:consym}
C^2(\rho_Q)=|c_A|^2+2\Delta\left(f_1- c_A d_2\right)
+\Delta^2(2|d_1|^2-c_A d_3-c_A^\ast d_3^\ast)\\
-2\Delta\Big[f_1^2+|c_A|^2 h - (c_A^\ast k + c_A k^\ast)+\Delta^2|d_1^2 - c_A^\ast d_3|^2\\
-2\Delta\left(|d_1|^2f_1+c_A c^\ast_A g-c_A d_1f_2-c_A^\ast d_1^\ast f_2^\ast\right)\Big]^{1/2},
\end{multline}
where
\begin{align}
d_1={}&\abra\mfw^\dagger\bket\label{eq:d1}\\
d_2={}&\int d\Pi_2\,\bbra\mathcal{M}^{\dagger}\mathcal{M}\aket\\
d_3={}&\abra\mfw^\dagger\mtilfw\bket\\
f_1={}&\int d\Pi_2\,\bbra\mathcal{M}\pa\mathcal{M}^\dagger\bket\\
f_2={}&\bbra\mtilfw^\dagger\int d\Pi_2(\mathcal{M}\pa\mathcal{M}^\dagger)\bket\\
g={}&\bbra\mtilfw^\dagger\int d\Pi_2(\mathcal{M}\pa\mathcal{M}^\dagger)\mtilfw\bket\\
h={}&\textrm{Tr}\big[\int d\Pi_2(\mathcal{M}\pa\mathcal{M}^\dagger)\int d\Pi_2(\mtil\prb\mtil^\dagger)\big]\\
k={}&\bbra\int d\Pi_2(\mathcal{M}\pa\mathcal{M}^\dagger)\int d\Pi_2(\mtil\prb\mtil^\dagger)\aket\,.\label{eq:k}
\end{align}

\paragraph{Initial product state}
Equation~(\ref{eq:consym}) simplifies enormously for an initial product state. In that case, in fact, $c_A=0$ 
and the concurrence squared is easily recast as
\be
C^2(\rho_Q)=2\,\Delta\, f_1+2\,\Delta^2\, |d_1|^2-2\,\Delta \sqrt{\left(f_1-\Delta\, |d_1|^2 \right)^2}=4\,\Delta^2\, |d_1|^2\,,
\ee
if $\Delta< f_1/|d_1|^2$\,.  Using \refeq{eq:d1} one finds
\be\label{eq:conpr}
C(\rho_{Q})=2\,\Delta\,| \bbra\mfw\aket|\simeq 2\,\Delta\, |\bbra\textrm{Re}\,\mfw\aket|\,,
\ee
at the leading order in perturbation theory. Equation~(\ref{eq:conpr}) shows that the entanglement of a pair of qubits is generated in scattering theory by the forward amplitude. In perturbation theory, if $\textrm{Re}\,\mfw\neq 0$, unitarity implies that the forward amplitude is dominated at the leading order by its real part. The contribution to the entanglement of the imaginary part or, equivalently, the scattering cross section, is instead suppressed by one perturbative order. 

It is convenient to rewrite \refeq{eq:conpr}, showing explicitly the qubit indices. Denoting the ``bar'' indices as $\bar{1}=2$ and $\bar{2}=1$, one has
\be
C(\rho_{Q})=2\,\Delta\left|\sum_{\alpha,\beta,\gamma,\delta}(-1)^{\alpha+\beta}a_{\alpha\beta}\,a_{\gamma\delta}(-\mfw)_{\bar{\alpha}\bar{\beta},\gamma\delta}\right|\,.
\ee
For an initial state belonging to the computational basis the amplitudes read $a_{\alpha\beta}=1$ and $a_{\alpha\bar{\beta}}=a_{\bar{\alpha}\beta}=a_{\bar{\alpha}\bar{\beta}}=0$, so that  $C(\rho_{Q})=2\Delta |(\mfw)_{\bar{\alpha}\bar{\beta},\alpha\beta}|$. In this case one can say that the concurrence is generated by the ``inelastic'' forward amplitude.\footnote{For a generic product state, one may be interested in finding the conditions that the elements of a scattering amplitude should satisfy in order to minimize qubit entanglement generated in a scattering process~(cf.~Refs.\cite{Carena:2023vjc,Chang:2024wrx,McGinnis:2025brt}). Given \refeq{eq:conpr}, in order to enforce $C(\rho_Q)=0$ the following conditions hold:
\bea\label{eq:cond}
&&\alpha=\gamma\wedge \beta=\delta\,:\,\sum_{\alpha,\beta}(-1)^{\alpha+\beta}a_{\alpha\beta}^2(-\mfw)_{\bar{\alpha}\bar{\beta},\alpha\beta}\implies (\mfw)_{\bar{\alpha}\bar{\beta},\alpha\beta}=0\,,\nonumber\\
&&\alpha\neq \gamma\wedge \beta\neq \delta\,:\,\sum_{\alpha,\beta}(-1)^{\alpha+\beta}a_{\alpha\beta}a_{\bar{\alpha}\bar{\beta}}(-\mfw)_{\bar{\alpha}\bar{\beta},\bar{\alpha}\bar{\beta}}\implies \sum_{\alpha,\beta}(-1)^{\alpha+\beta}(-\mfw)_{\bar{\alpha}\bar{\beta},\bar{\alpha}\bar{\beta}}=0\,,\nonumber\\
&&\alpha\neq \gamma \wedge \beta=\delta\,:\,\sum_{\alpha,\beta}(-1)^{\alpha+\beta}a_{\alpha\beta}a_{\bar{\alpha}\beta}(-\mfw)_{\bar{\alpha}\bar{\beta},\bar{\alpha}\beta}\implies \forall_{\beta}\sum_{\alpha}(-1)^{\alpha+\beta}(-\mfw)_{\bar{\alpha}\bar{\beta},\bar{\alpha}\beta}=0\,,\nonumber\\
&&\alpha=\gamma \wedge \beta\neq \delta\,:\,\sum_{\alpha,\beta}(-1)^{\alpha+\beta}a_{\alpha\beta}a_{\alpha\bar{\beta}}(-\mfw)_{\bar{\alpha}\bar{\beta},\alpha\bar{\beta}}\implies \forall_{\alpha}\sum_{\beta}(-1)^{\alpha+\beta}(-\mfw)_{\bar{\alpha}\bar{\beta},\alpha\bar{\beta}}=0\,.
\eea
Writing the requirements of Eqs.~(\ref{eq:cond}) explicitly, one identifies relations between the elements of the forward scattering amplitude:
\begin{align}\label{eq:minent}
(\mfw)_{22,11}=(\mfw)_{21,12}&=(\mfw)_{12,21}=(\mfw)_{11,22}=0\,,\nonumber\\
(\mfw)_{21,21}+(\mfw)_{12,12}&=(\mfw)_{11,11}+(\mfw)_{22,22}\,,\nonumber\\
(\mfw)_{11,21}=(\mfw)_{12,22},&\qquad (\mfw)_{11,12}=(\mfw)_{21,22}\,.
\end{align}
The same relations were previously pointed out in Refs.\cite{Carena:2023vjc,Chang:2024wrx,McGinnis:2025brt}. }

Note that \refeq{eq:conpr} applies to the case of initial states with highly collimated momentum distribution, approximated by an element of the basis, $|\mathbf{p}_A\mathbf{p}_B\rangle$. In light of the discussion in  
Appendix~\ref{app:WP} it is easy to infer the form of the concurrence $C(\rho_{Q})$ when the wave packet is not peaked:
\be
C(\rho_{Q})=2 |\bbra\textrm{Re}\,\mathcal{M}_{\textrm{overlap}}\aket|\,,
\ee
where 
\be
\mathcal{M}_{\textrm{overlap}}\equiv \frac{1}{16\pi}\int_{\textrm{supp}\,\phi_{A,B}} d(\cos\theta_i)\, \overline{\mathcal{M}}(p_A,p_B,\theta_i)\,,
\ee
defined in \refeq{eq:ot_wave_up} of Appendix~\ref{app:WP}, quantifies the solid-angle overlap of the final-state momentum distribution with the initial state.

A physical interpretation of \refeq{eq:conpr} would be that the dominant contribution to qubit entanglement is generated by quantum correlations between the unscattered wave along the beamline and the plane wave that sees its quantum numbers changed by the interaction yet maintains a nearly unaltered momentum. At higher orders in the perturbative expansion one still expects entanglement to be generated between two particles at all values of the angular final-state distribution. However, this will be perturbatively suppressed with respect to the forward contribution if the latter exists.

\section{Applications\label{sec:examp}}

In this section, we show two practical applications of the properties we have derived for the linearized entropy in \refsec{sec:entpow}, and for the  concurrence in \refsec{sec:conc}. 

\subsection{The 2HDM\label{sec:2HDM}}

The first model we analyze is the 2HDM. The 2HDM was recently 
used as a tool to investigate the potential emergence of global symmetries of the Lagrangian from the minimization of qubit entanglement in the scattering process\cite{Carena:2023vjc}. It was shown in Refs.\cite{Kowalska:2024kbs,Chang:2024wrx} that, once all of the scattering processes allowed by the gauge symmetry were considered, the emergence of an additional global symmetry could not be established. Despite this negative result, it is instructive to recast some of the quantities derived in Ref.\cite{Kowalska:2024kbs} in the light of \refsec{sec:entpow} and
\refsec{sec:conc}, so to extract some generic properties of qubit entanglement in this model.

The 2HDM consists of two scalar doublets of the $SU(2)_L$ group of the SM, $H_1$ and $H_2$, which carry  
hypercharge value~$1/2$. Assuming there exists some self-adjoint observable allowing one to distinguish the two doublets after the scattering process, one can construct the qubit out of the doublet's index, $\alpha=1,2$. Following Ref.\cite{Carena:2023vjc}, we dub this index as the field's ``flavor.'' 

The gauge-invariant scalar potential of the model is given by\cite{Botella:1994cs}
\begin{multline}\label{eq:scapot0}
V(H_1,H_2)=\mu_1^2\,H_1^\dag H_1+\mu_2^2\,H_2^\dag H_2+\left(\mu_3^2\, H_1^\dag H_2+\textrm{H.c.}\right)\\
+\lambda_1\,(H_1^\dag H_1)^2+\lambda_2\,(H_2^\dag H_2)^2+\lambda_3\,(H_1^\dag H_1)(H_2^\dag H_2)+\lambda_4\,(H_1^\dag H_2)(H_2^\dag H_1)\\
+\left(\lambda_5\,(H_1^\dag H_2)^2+\lambda_6\,(H_1^\dag H_1)(H_1^\dag H_2)+\lambda_7\,(H_2^\dag H_2)(H_1^\dag H_2)+\textrm{H.c.}\right),
\end{multline}
where $\mu^2_{1,2,3}\geq 0$ are the squared mass parameters, which we take to be positive, 
and $\lambda_{1,\dots,7}$ denote the dimensionless quartic couplings, which we assume to be real.  The scalar doublets can be explicitly decomposed into charged and neutral components, $H_\alpha=[\,h_\alpha^+,h_\alpha^0\,]^T$.

We are interested in the scattering of two complex 
fields, $h_\alpha\,h_\beta\to h_\gamma\,h_\delta$,
at high energy ($s\gg \mu^2_{1,2,3}$). In the ultrarelativistic regime this is dominated by contact interactions. The gauge symmetry of the scalar potential (\ref{eq:scapot0}) allows for five different scattering processes.  The amplitude matrices $\mathcal{M}^{(n)}_{\alpha\beta,\gamma\delta}$ can be found -- expressed at the $n=1$ 
loop level in the flavor basis $|11\rangle$, $|12\rangle$, $|21\rangle$, $|22\rangle$ -- in Eqs.~(3.2)-(3.12) of Ref.\cite{Kowalska:2024kbs}.

The amplitude presents 
the same texture in all five combinations of neutral and charged scalar scattering processes enlisted in Ref.\cite{Kowalska:2024kbs}. In order to derive the parametric dependence of the amplitude we focus here on one particular process, $h^0 h^0\to h^0 h^0$. Results relative to the other four processes can be derived straightforwardly with an index substitution. 
We recall the tree-level scattering matrix,
\be\label{eq:scat00}
i\mathcal{M}^{(0)}(h^0h^0\to h^0h^0)=
-i\left(\begin{array}{cccc} 
4\lambda_1 & 2\lambda_6 & 2\lambda_6 & 4\lam_5\\ 
2\lambda_6 & \lambda_3+\lambda_4 & \lambda_3+\lambda_4 & 2\lambda_7 \\
2\lambda_6 & \lambda_3+\lambda_4 & \lambda_3+\lambda_4 & 2\lambda_7 \\
4\lam_5 & 2\lambda_7 & 2\lambda_7 & 4\lambda_2
\end{array}\right)\,,
\ee
while the $\overline{MS}$ 1-loop level elements can be found in Eqs.~(3.7)-(3.12) of Ref.\cite{Kowalska:2024kbs}.

\paragraph{Product state} For practical purposes, we shall consider 
an element of the computational basis in the initial state, $|A\rangle=|11\rangle$, so that
$\mathcal{P}_{A,11}\neq 0$ and all other entries of the projector~$\pa$ are zero. 
Applying \refeq{rho:flav} one computes the 1-loop density matrix\cite{Kowalska:2024kbs}, 
\be\label{eq:rho11Ex}
\rho_{Q(11,11)}=1-\Delta\left(\frac{\lam_5^2}{\pi}+\frac{\lam_6^2}{2\pi}\right)\,,
\ee
\be
\rho_{Q(11,12)}=\rho_{Q(11,21)}=\rho_{Q(12,11)}^{\ast}=\rho_{(21,11)}^{\ast}=
\Delta\left( 2\,i\,\lam_6 +\frac{2\lam_1 \lam_6-\lam_3\lam_6-\lam_4 \lam_6-2\lam_5 \lam_7 }{8\pi} \right)\,,
\ee
\be
\rho_{Q(11,22)}=\rho_{Q(22,11)}^{\ast}=
\Delta\left( 4\,i\,\lam_5 +\frac{2\lam_1 \lam_5-2\lam_2\lam_5-\lam_6\lam_7 }{4\pi} \right)\,,
\ee
\be
\rho_{Q(12,12)}=\rho_{Q(12,21)}=\rho_{Q(21,12)}^{\ast}=\rho_{Q(21,21)}=\Delta\,\frac{\lam_6^2}{4\pi}\,,
\ee
\be
\rho_{Q(12,22)}=\rho_{Q(21,22)}=\rho_{Q(22,12)}^{\ast}=\rho_{Q(22,21)}^{\ast}=\Delta\,\frac{\lam_5 \lam_6}{2\pi}\,,
\ee
\be\label{eq:rho22Ex}
\rho_{Q(22,22)}=\Delta\,\frac{\lam_5^2}{\pi}\,.
\ee

At the leading order, $\rho_Q$ features two non-zero eigenvalues that read
\bea
e_1&=&\left(\lam_5^2+\frac{\lam_6^2}{2} \right)\left(\frac{\Delta}{\pi}-16\,\Delta^2 \right),\label{eq:eig1}\\
e_2&=&1-\left(\lam_5^2+\frac{\lam_6^2}{2} \right)\left(\frac{\Delta}{\pi}-16\,\Delta^2 \right).\label{eq:eig2}
\eea
The corresponding linearized entropy reads 
\be\label{eq:lin2HDM}
\mathcal{E}\equiv 1-\textrm{Tr}\rho_Q^2=\left(2\lam_5^2+\lam_6^2 \right)\left(\frac{\Delta}{\pi}-16\,\Delta^2\right),
\ee
which leads to a unitarity bound, $\Delta\leq 1/(16 \pi)$. 

The total and ``elastic'' cross section from contact interactions can be easily computed at the tree level,
\bea
\sigma_{|11\rangle\to\textrm{all}}&=&\frac{1}{4\pi s}\left(2\lam_1^2+2\lam_5^2+\lam_6^2 \right),\\
\sigma_{|11\rangle\to |11\rangle} &=& \frac{\lam_1^2}{2\pi s}\,.
\eea
These expressions confirm the area law~(\ref{eq:arealaw}) at order $\Delta$. Note, however,
that the linearized entropy in \refeq{eq:linen} presents also a correction due to the real part of the forward amplitude which, for isotropic scattering, coincides with 
the integrated amplitude. One can thus compute 
\bea
2\, \langle 11| \mfw^{\dag}\mfw |11\rangle & = & 16\left(2\lam_1^2+2\lam_5^2+\lam_6^2 \right),\\
\langle 11| \mfw \pa \mfw | 11\rangle & = &  16\,\lam_1^2\,,\\
\langle 11| \mfw^{\dag} \pa \mfw^{\dag} | 11\rangle & = & 16\,\lam_1^2\,,
\eea
which correct the area law and show that \refeq{eq:lin2HDM} agrees with \refeq{eq:linen}. 

We extract the concurrence from the square roots of the eigenvalues of matrix $R=\rho_Q \tilde{\rho}_Q$\,. There are two non-zero eigenvalues, 
\bea
\xi_1&=&\frac{\lam_5^2\,\Delta}{\pi}+16\,\lam_5^2\,\Delta^2-\sqrt{\frac{64\,\lam_5^4\,\Delta^3}{\pi}}\,,\label{eq:lam1}\\
\xi_2&=&\frac{\lam_5^2\,\Delta}{\pi}+16\,\lam_5^2\,\Delta^2+\sqrt{\frac{64\,\lam_5^4\,\Delta^3}{\pi}}\,,\label{eq:lam2}
\eea
which lead to\footnote{Equation~(\ref{eq:C2HDM}) rightly features the leading $\Delta$ dependence of the concurrence as $\Delta^1$, correcting an error in the equivalent expression of Ref.\cite{Kowalska:2024kbs}.}
\be\label{eq:C2HDM}
C(\rho_Q)=|\sqrt{\xi_1}-\sqrt{\xi_2}| = 8\,\Delta\, \lam_5+\mathcal{O}(\lam^2)\,.
\ee

To make contact with the results in \refsec{sec:conc} we take 
$|A\rangle=|11\rangle$ and $|B\rangle=-|22\rangle$. One finds straightforwardly
$|\langle B|\mfw|A\rangle | = 4\,\lam_5$ at the leading order. Equation~(\ref{eq:C2HDM}) thus agrees with \refeq{eq:conpr} in the 2HDM.

\paragraph{Bell state} Let us now consider 
a maximally entangled Bell state, 
\be
|A\rangle=\frac{1}{\sqrt{2}}\left(|11\rangle+|22\rangle\right).
\ee

From \refeq{rho:flav}
one can construct the density matrix at one loop:
\be\label{eq:rhoB1Ex}
\rho_{Q(11,11)}=\frac{1}{2}+\frac{\Delta}{4\pi}\left(2\lam_1 \lam_5-2\lam_2\lam_5-\lam_6^2-\lam_6 \lam_7 \right),
\ee
\begin{multline}
\rho_{Q(11,12)}=\rho_{Q(11,21)}=\rho_{Q(12,11)}^{\ast}=\rho_{(21,11)}^{\ast}
=\Delta\left[i\left(\lam_6+\lam_7 \right)\right.\\
\left.+\,\frac{1}{16\pi}\left(2\lam_1 \lam_6-\lam_3\lam_6 -\lam_4 \lam_6+2\lam_5 \lam_6 + 4\lam_1 \lam_7-2\lam_2 \lam_7-\lam_3 \lam_7-\lam_4 \lam_7+2\lam_5\lam_7\right) \right],
\end{multline}
\be
\rho_{Q(11,22)}=\rho_{Q(22,11)}^{\ast}=\frac{1}{2}-\Delta\left[ 2 i \left(\lam_1-\lam_2\right)+\frac{2\lam_1^2-4\lam_1\lam_2+2\lam_2^2+\lam_6^2
+2\lam_6 \lam_7+\lam_7^2}{8\pi}\right],
\ee
\be
\rho_{Q(12,12)}=\rho_{Q(12,21)}=\rho_{Q(21,12)}^{\ast}=\rho_{Q(21,21)}=\Delta\,\frac{\left(\lam_6+\lam_7\right)^2}{8\pi}\,,
\ee
\begin{multline}
\rho_{Q(12,22)}=\rho_{Q(21,22)}=\rho_{Q(22,12)}^{\ast}=\rho_{Q(22,21)}^{\ast}
=\Delta\left[-i\left(\lam_6+\lam_7 \right)\right.\\
\left.-\,\frac{1}{16\pi}\left(2\lam_1 \lam_6-4\lam_2 \lam_6+\lam_3\lam_6+\lam_4\lam_6
-2\lam_5\lam_6-2\lam_2 \lam_7+\lam_3\lam_7+\lam_4\lam_7-2\lam_5 \lam_7 \right) \right],
\end{multline}
\be\label{eq:rho22Exx}
\rho_{Q(22,22)}=\frac{1}{2}+\frac{\Delta}{4\pi}\left(-2\lam_1 \lam_5+2\lam_2\lam_5-\lam_7^2-\lam_6 \lam_7 \right).
\ee

Following the same steps as above one finds two non-zero eigenvalues at the leading order,
\bea
e_1&=&\left(\frac{\Delta}{4\pi}-4\Delta^2 \right)\left[\left(\lam_1-\lam_2\right)^2+\left(\lam_6+\lam_7\right)^2\right],\\
e_2&=&1-\left(\frac{\Delta}{4\pi}-4\Delta^2 \right)\left[\left(\lam_1-\lam_2\right)^2+\left(\lam_6+\lam_7\right)^2\right],
\eea
and the linearized entropy,
\be
\mathcal{E}=\left(\frac{\Delta}{2\pi}-8\Delta^2 \right)\left[\left(\lam_1-\lam_2\right)^2+\left(\lam_6+\lam_7\right)^2\right].
\ee

We extract the concurrence from the traces of matrix $R=\rho_Q \tilde{\rho}_Q$\,, following 
\refeq{eq:contr}. One finds in this case,\footnote{Equation~(\ref{eq:conBell2HDM}) corrects a factor-of-2 error in Eq.~(3.32) of our previous paper\cite{Kowalska:2024kbs}.}
\be\label{eq:conBell2HDM}
C(\rho_Q)= 1-\left(\frac{\Delta}{2\pi}-8\Delta^2 \right)\left[\left(\lam_1-\lam_2\right)^2+\left(\lam_6+\lam_7\right)^2\right],
\ee
which implies 
\be\label{eq:conBell}
C = 1-\mathcal{E}
\ee
for the Bell state, at the leading order in perturbation theory.

As was first observed in Ref.\cite{Kowalska:2024kbs} for the 2HDM,
entanglement seems to ``flow'' between different partitions of the Hilbert space. 
Equation~(\ref{eq:conBell}) gives a quantitative parametrization of the flow.
Suppose that one starts 
with an initial separable $|\textrm{in}\rangle$ state in $\mathcal{H}_{\textrm{mom}}\otimes \mathcal{H}_{\textrm{qb}}$, a 
state that is the product of a momentum wave function and a Bell state and thus features
$\mathcal{E}_{\textrm{in}}=0$ and $C_{\textrm{in}}=1$. Interestingly, after the scattering process a certain amount of entanglement is transferred from the qubit Bell state to the product state of momentum and flavor as described in \refeq{eq:conBell}:
$C_{\textrm{out}}= 1-\mathcal{E}_{\textrm{out}}$, 
with $\mathcal{E}_{\textrm{out}}\neq 0$. It is an intriguing question whether 
\refeq{eq:conBell} is a property of contact interactions only, or it is a more generic feature of Bell states in perturbative scattering theory. The answer to this question is not trivial, and cannot be straightforwardly evinced from \refeq{eq:consym}. We thus leave a detailed investigation of this issue for future work. 

\subsection{$e^+e^-$ annihilation\label{sec:QED}}

Let us now consider a situation in which the real part of the inelastic forward amplitude is missing due to an underlying symmetry. 
This is the case, for example, of $e^+ e^-\to \mu^+ \mu^-$ scattering in QED at very high energies, $s\gg m_{\mu}^2$\,. 

Entanglement in QED was analyzed in a number of papers\cite{Cervera-Lierta:2017tdt,Fonseca:2021uhd,Ghodrati:2023uef,Blasone:2024dud,Blasone:2024jzv,Blasone:2025tor,Blasone:2025ddi,Martin:2025hzm}. 
The $e^+ e^-\to \mu^+ \mu^-$ scattering process is mediated by the exchange of a photon in the $s$-channel and the 
amplitude is well known:
\be
i\mathcal{M}_{s}(p_a,p_b\to p_i,p_j)=\frac{i e^2}{s\left[1-\Pi(q^2)\right]}\left[\bar{v}(p_b)\gamma^{\mu} u(p_a) \right] \left[\bar{u}(p_i)\gamma_{\mu} v(p_j) \right],
\ee
where $\Pi(q^2)$ gives the non-tensorial part of 1PI insertions to the photon propagator, $p_a(e^-)$, $p_b(e^+)$ are initial momenta, 
$p_i(e^-)$, $p_j(e^+)$ are final momenta, $s=(p_a+p_b)^2$, and $e$ is the coupling constant. The imaginary part of $\Pi(0)$, which can be computed from the vacuum polarization diagram at 
one loop, reads\cite{Peskin:1995ev}
\be
\textrm{Im}\,\Pi^{(1)}(0)=-\frac{e^2}{12\pi}\,.
\ee

Following common practice, 
we construct the qubit out of the two eigenvectors 
of the helicity operator, $\boldsymbol{\sigma}\cdot\mathbf{p}$\,. 
The corresponding index spans the values $i=L,R$, and  
one can express the amplitude in a computational basis built out of the helicity states of the scattering fermions:
$|LL\rangle$, $|LR\rangle$, $|RL\rangle$, $|RR\rangle$. In the high-energy limit 
it reads 
\be\label{eq:MQED}
\mathcal{M}(\theta)=e^2\left(1-\frac{i\,e^2}{12\pi}\right)
\left(\begin{array}{cccc} 
0 & 0 & 0 & 0\\ 
0 & -1-\cos\theta & 1-\cos\theta & 0 \\
0 & 1-\cos\theta & -1-\cos\theta & 0 \\
0 & 0 & 0 & 0
\end{array}\right)\,,
\ee
where $\theta$ is the scattering angle with respect to the incoming ($z$) direction. As is well known, the zero elements of matrix~$\mathcal{M}$ are due to the insertion of helicity-projection operators $P_L$, $P_R$ between the spinors. Moreover, two additional zeros emerge in the forward direction, at $\theta=0$, due to the conservation of total angular momentum\cite{Peskin:1995ev}.

Let us consider an initial state belonging to the computational basis, $|A\rangle=|LR\rangle$, so that
$\mathcal{P}_{A,LR}\neq 0$ and all other entries of the projector $\mathcal{P}_A$ are zero. Unlike in \refsec{sec:2HDM}, this will be the only case we discuss, as the form of \refeq{eq:MQED} makes Bell states trivially invariant under the $S$-matrix.
Applying \refeq{rho:flav} one finds that at one loop  
the density matrix of the final state acquires a simple form,
\be\label{eq:rhoQED}
\rho_Q=
\left(\begin{array}{cccc} 
0 & 0 & 0 & 0\\ 
0 & 1-\frac{e^4 \Delta}{6\pi} & -\frac{e^4 \Delta}{12\pi} & 0 \\
0 & -\frac{e^4 \Delta}{12\pi} & \frac{e^4 \Delta}{6\pi} & 0 \\
0 & 0 & 0 & 0
\end{array}\right)\,.
\ee

At the leading order, the linearized entropy is given by
\be\label{eq:linQED}
\mathcal{E}=1-\textrm{Tr}\rho_Q^2= \frac{e^4 \Delta}{3\pi}\,.
\ee
One can check that the area law~(\ref{eq:arealaw}) holds, 
as can be evinced from the well-known expressions of the total and ``elastic'' cross section: 
\bea
\sigma_{|LR\rangle\to\textrm{all}}&=&\frac{e^4}{6\pi s}\,,\\
\sigma_{|LR\rangle\to |LR\rangle} &=& \frac{e^4}{12\pi s}\,.
\eea

On the other hand, a comparison with \refeq{eq:linen} shows that the $\mathcal{O}(\Delta^2)$ term is absent in \refeq{eq:linQED}. This is consistent with the form of the forward amplitude, which yields
\bea
2\, \langle LR| \mfw^{\dag}\mfw |LR\rangle & = & 8e^4\,,\\
\langle LR| \mfw \pa \mfw | LR\rangle & = & 4e^4\,,\\
\langle LR| \mfw^{\dag} \pa \mfw^{\dag} | LR\rangle & = & 4e^4\,,
\eea
and thus cancels out of \refeq{eq:linQED}.

The concurrence shows as well an important difference with the example in \refsec{sec:2HDM}. 
Let us compute the square roots of the eigenvalues of matrix $R=\rho_Q \tilde{\rho}_Q$\,. There are two non-zero eigenvalues, which at the leading order read
\be
\xi_1 \simeq \xi_2 = \frac{e^4 \Delta}{6\pi}+\mathcal{O}(e^8)\,.
\ee
One can see that their difference will not contribute to the concurrence at the lowest order in perturbation theory. 
Contributions of the order of $|\mathcal{M}|^2/16\pi$ will be generated, but one will need a 2-loop analysis of the amplitude to determine their exact expression. This finding agrees with 
\refeq{eq:conpr}. Given, in fact, $|A\rangle=|LR\rangle$ and $|B\rangle=|RL\rangle$, one finds straightforwardly
\be
\langle B|\mfw |A\rangle = e^2\left(1-\cos\theta \right)\left|_{\theta=0}\right. = 0\,. 
\ee

\section{Summary and conclusions\label{sec:sum}}

In this work, we derived an analytic expression for the concurrence of two qubits in the mixed final state of a relativistic $2\to 2$ scattering event. We relied on a perturbative expansion of the $S$-matrix, which led to the concurrence being expressed in terms of the scattering amplitude and initial state only. We found that, at the leading order, 
the concurrence depends exclusively on the real part of the ``inelastic'' forward scattering amplitude, in contrast to the linearized entropy of the reduced density matrix, which depends instead on its imaginary part or, equivalently, the total cross section. 

Provided one starts with a product state, our formula allows to readily compute the dominant contribution 
to the entanglement generated by the scattering event bypassing the construction of the density matrix of the (generally mixed) final state. Moreover, unitarity is fully preserved.  
In a complementary result, we also showed that the real part of the forward amplitude provides a subleading correction to the linearized entropy, which reduces the latter by an amount connected to the relative entropy of coherence of the final state after scattering. 
  
For illustration, we have applied our results to two cases of phenomenological interest. First, the high-energy scattering of two ``flavored'' scalar fields in the 2HDM, where, by the very nature of contact interactions, the amplitude is isotropic and its forward part is the same as at all other scattering angles. Second, the annihilation of $e^+ e^-$ helicity states in QED at high energy, where, due to the conservation of total angular momentum, the inelastic part of the amplitude is absent in the forward direction, but exists at higher $p_T$. 

An extension to quantum states with more than one discrete quantum number (one could think, for example, of a 2HDM situation where, besides the flavor qubit, the electric charge is promoted to a qutrit) is straightforward at the level of the density matrix, as it simply implies the tracing-out of different partitions. On the other hand, the derivation of a concurrence expression that depends only on the scattering amplitude appears to be significantly more challenging in the presence of more elaborated partitions. It may provide a cue for future investigations. 

Another interesting direction for forthcoming investigations would be the study of how our results translate to the partial-wave and phase shifts formalism, which may possibly bypass the treatment of indeterminate quantities like the parameter $\Delta$.  

Finally, our concurrence result implies that the dominant contribution to qubit entanglement appears to be generated by quantum correlations between the unscattered particles and particles that interact but see their momentum remain nearly unchanged. Needless to say, at higher orders in the perturbative expansion one expects entanglement to be generated between two scattered particles at all values of the angular final-state distribution, yet this will be perturbatively suppressed with respect to the forward contribution if the latter exists. Thus, in the context of possible emergent symmetries from an entanglement extremization principle, one may envision a situation where a symmetry is identified at the leading order from the conditions of vanishing concurrence, yet it is later spoiled by higher-order effects. 
This possibly suggests that a non-perturbative approach to the $S$-matrix may help to settle the issue of emergent symmetries in a definitive way. 

\bigskip
 \begin{center}
 \textbf{ACKNOWLEDGMENTS}
 \end{center}
EMS is supported in part by the National Science Centre (Poland) under the research Grant No.~2020/38/E/ST2/00126. 
\bigskip

\appendix
\addcontentsline{toc}{section}{Appendices}


\section{Generic wave packets\label{app:WP}}

In this appendix, we generalize \refeq{eq:ot} to the case of 
arbitrary wave packets~$\phi^{\alpha}_{A,B}(\mathbf{p})$ with $L^2(\mathbb{R}^3)\otimes \mathbb{C}^d$ norm 
\be
||\phi_{A,B}||^2=\sum_{\alpha=1}^d\int \frac{d^3 p}{(2\pi)^3}\left| \phi^{\alpha}_{A,B}(\mathbf{p}) \right|^2 = 1\,.
\ee

One can write \refeq{eq:ot} explicitly as 
\begin{multline}\label{eq:ot_wave}
\langle\textrm{out}|\textrm{out}\rangle=1+ \left[ \sum_{\alpha\beta,\gamma\delta} \int\int\frac{d^3 p_i}{(2\pi)^3}\frac{1}{\sqrt{2 E_i}}\frac{d^3 p_j}{(2\pi)^3}\frac{1}{\sqrt{2 E_j}}
\int \int \frac{d^3 p_a}{(2\pi)^3}\frac{1}{\sqrt{2 E_a}}\frac{d^3 p_b}{(2\pi)^3}\frac{1}{\sqrt{2 E_b}}\right.\\
\left. (2\pi)^4 \delta^4 (p_a+p_b-p_i-p_j)
 \phi_{A}^{\gamma\ast}(\mathbf{p}_i) \phi_{B}^{\delta\ast}(\mathbf{p}_j) i \mathcal{M}_{\gamma\delta,\alpha\beta}(p_a,p_b\to p_i,p_j) \phi_{A}^{\alpha}(\mathbf{p}_a) \phi_{B}^{\beta}(\mathbf{p}_b) +\textrm{c.c.}\right]\\
+ \sum_{\gamma\delta,\alpha\beta,\epsilon\sigma} \int\int\frac{d^3 p_i d^3 p_j}{(2\pi)^6 4 E_i E_j}
\int\int\int\int \frac{d^3 p_a d^3 p_b d^3 p_c d^3 p_d}{(2\pi)^6\sqrt{4 E_a E_b}(2\pi)^6 \sqrt{4 E_c E_d}}\\
\times (2\pi)^8 \delta^4(p_a+p_b-p_i-p_j) \delta^4(p_c+p_d-p_i-p_j)\\
\phi_A^{\epsilon\ast}(\mathbf{p}_c)\phi_B^{\sigma\ast}(\mathbf{p}_d) 
\mathcal{M}^{\ast}_{\gamma\delta,\epsilon\sigma}(p_c,p_d\to p_i, p_j)
\mathcal{M}_{\gamma\delta,\alpha\beta}(p_a,p_b\to p_i, p_j)
\phi_A^{\alpha}(\mathbf{p}_a)\phi_B^{\beta}(\mathbf{p}_b)\,.
\end{multline}

Let us focus first on the second addend in \refeq{eq:ot_wave} and work in the c.o.m.~frame.
One can start by eliminating the 4-dimensional delta function and compute
\begin{multline}\label{eq:ot_wave_up}
\langle\textrm{out}|\textrm{out}\rangle=1+\left[ \sum_{\alpha\beta,\gamma\delta} \int\frac{d^3 p_i}{(2\pi)^2}\frac{1}{\sqrt{4 E_i E_j}} \delta (\sqrt{s}-E_i-E_j)
\int \int \frac{d^3 p_a d^3 p_b}{(2\pi)^6}\frac{1}{\sqrt{4 E_a E_b}}\right.\\
\left. \phi_{A}^{\gamma\ast}(\mathbf{p}_i) \phi_{B}^{\delta\ast}(-\mathbf{p}_i) i \mathcal{M}_{\gamma\delta,\alpha\beta}(p_a,p_b\to p_i,\tilde{p}_i) \phi_{A}^{\alpha}(\mathbf{p}_a) \phi_{B}^{\beta}(\mathbf{p}_b) +\textrm{c.c.}\right]...\\
 = 1+\left[i  \sum_{\alpha\beta,\gamma\delta} \int\int\frac{|\mathbf{p}_i(X)|d X d \Omega_i}{4\,(2\pi)^2} \delta (\sqrt{s}-X)
\int \int \frac{d^3 p_a d^3 p_b}{(2\pi)^6 \sqrt{s}}\right.\\
\left. \phi_{A}^{\gamma\ast}(\mathbf{p}_i) \phi_{B}^{\delta\ast}(-\mathbf{p}_i) \mathcal{M}_{\gamma\delta,\alpha\beta}(p_a,p_b\to p_i,\tilde{p}_j) \phi_{A}^{\alpha}(\mathbf{p}_a) \phi_{B}^{\beta}(\mathbf{p}_b) +\textrm{c.c.}\right]...\\
 \simeq 1+\left[\frac{i}{16\pi}\int_{\textrm{supp}\,\phi_{A,B}} d(\cos\theta_i)\, \overline{\mathcal{M}}(p_A,p_B;\theta_i) +\textrm{c.c.}\right]...\,,
\end{multline}
where we have defined $X=E_i+E_j$ and 
used the identity (\textit{e.g.},\cite{Halzen:1984mc})
\be
d p_i=\frac{dX}{|\mathbf{p}_i(X)|}\frac{E_i E_j}{X}\,,
\ee
while sending $2\,|\mathbf{p}_i|/\sqrt{s}\to 1$ in the high-energy limit. 
In the last step we have defined 
\be\label{eq:mtil}
\overline{\mathcal{M}}(p_A,p_B;\theta_i)= \sum_{\alpha\beta,\gamma\delta} \int \int \frac{d^3 p_a d^3 p_b}{(2\pi)^6}
 \phi_{A}^{\gamma\ast}(\theta_i) \phi_{B}^{\delta\ast}(\theta_i) \mathcal{M}_{\gamma\delta,\alpha\beta}(p_a,p_b;\theta_i) \phi_{A}^{\alpha}(\mathbf{p}_a) \phi_{B}^{\beta}(\mathbf{p}_b)\,,
\ee
and computed the $\cos\theta_i$ integral over the support of the 
initial-state wave packets $\phi_{A,B}$. 

Equation~(\ref{eq:mtil}) means that the indeterminate parameter $\Delta$ defined in \refsec{sec:entpow} quantifies the solid-angle overlap of the final-state momentum distribution with the initial state. As a matter of fact, one could write 
the second addend in \refeq{eq:ot_wave} simply as 
\begin{multline}\label{eq:ot_wave_mod}
\langle\textrm{out}|\textrm{out}\rangle\\
=1+\left[\frac{i\,Z(g^2)\,g^2}{16\pi}\sum_{\gamma=1}^d\int\frac{d^3 p_i}{(2\pi)^3}\phi_A^{\gamma\ast}(\mathbf{p}_i)\tilde{\phi}_A^{\gamma}(\mathbf{p}_i)
\sum_{\delta=1}^d\int\frac{d^3 p_j}{(2\pi)^3}\phi_B^{\delta\ast}(\mathbf{p}_j)\tilde{\phi}_B^{\delta}(\mathbf{p}_j)
+\textrm{c.c.}\right]...\,,
\end{multline}
where $Z(g^2)\,g^2$ is the renormalized coupling constant and 
\be
Z(g^2) g^2\,\tilde{\phi}_A^{\gamma}(\mathbf{p}_i) \tilde{\phi}_B^{\delta}(\mathbf{p}_j) =\sum_{\alpha\beta} \int d(\cos\theta_a)\,\mathcal{M}_{\gamma\delta,\alpha\beta}(\theta_a; p_i,p_j)  \phi_{A}^{\alpha}(\theta_a) \phi_{B}^{\beta}(\theta_a)\,,
\ee
expressed in terms of the angular distribution of the initial wave packet $\phi^{\alpha,\beta}_{A,B}(\mathbf{p})$. 
In~\refeq{eq:ot_wave_mod}, one can apply Schwarz inequality: 
\be
|\langle \phi,\tilde{\phi}\rangle|\leq ||\phi||\, ||\tilde{\phi}||
\ee
for any pair $\phi$, $\tilde{\phi}$ of elements of the Hilbert space -- thus
proving that the last two integrals typically yield a complex number in modulus smaller than one. The integrals are equal to one if and only if $\tilde{\phi}^{\gamma,\delta}_{A,B}(\mathbf{p})=\phi^{\gamma,\delta}_{A,B}(\mathbf{p})$, a case that corresponds to $\Delta=\Delta_{\textrm{max}}$.

In order to determine $\Delta_{\textrm{max}}$ one ought to look at the third addend in \refeq{eq:ot_wave}. Following similar steps as above one finds
\begin{multline}\label{eq:amplsqua}
\sum_{\alpha\beta,\gamma\delta,\epsilon\sigma} \int\int\frac{|\mathbf{p}_i(X)| dX d\Omega_i}{4\,(2\pi)^2 X}  \delta(\sqrt{s}-X)
\int\int\int\int \frac{d^3 p_a d^3 p_b |\mathbf{p}_c(Y)| dY d\Omega_c }{(2\pi)^6\sqrt{s}\,4\, (2\pi)^2}  \delta(Y-X) \\
\phi_A^{\epsilon\ast}(\mathbf{p}_c) \phi_B^{\sigma\ast}(-\mathbf{p}_c) \mathcal{M}^{\ast}_{\gamma\delta,\epsilon\sigma}(p_c,\tilde{p}_c\to p_i, \tilde{p}_i)
\mathcal{M}_{\gamma\delta,\alpha\beta}(p_a,p_b\to p_i, \tilde{p}_i)\phi_A^{\alpha}(\mathbf{p}_a)
\phi_B^{\beta}(\mathbf{p}_b)\\
=\sum_{\alpha\beta,\gamma\delta,\epsilon\sigma} \int\frac{|\mathbf{p}_i(\sqrt{s})| d\Omega_i}{4\,(2\pi)^2 \sqrt{s}}
\int\int\int \frac{d^3 p_a d^3 p_b |\mathbf{p}_c(\sqrt{s})|  d\Omega_c }{(2\pi)^6\sqrt{s}\,4\, (2\pi)^2}\\
\phi_A^{\epsilon\ast}(\theta_c) \phi_B^{\sigma\ast}(\theta_c) \mathcal{M}^{\ast}_{\gamma\delta,\epsilon\sigma}(\theta_c;\theta_i)
\mathcal{M}_{\gamma\delta,\alpha\beta}(p_a,p_b;\theta_i)\phi_A^{\alpha}(\mathbf{p}_a)
\phi_B^{\beta}(\mathbf{p}_b)\\
\simeq \left(\frac{1}{16\pi} \right)^2 Z(g^2) g^2 \int_{\textrm{supp}\,\tilde{\phi}_{A,B}} d(\cos\theta_i)\overline{\overline{\mathcal{M}}}(p_A,p_B;\theta_i)\,,
\end{multline}
where 
\be
\overline{\overline{\mathcal{M}}}(p_A,p_B;\theta_i)=
\sum_{\alpha\beta,\gamma\delta} \int \int \frac{d^3 p_a d^3 p_b}{(2\pi)^6}
\tilde{\phi}_{A}^{\gamma\ast}(\theta_i) \tilde{\phi}_{B}^{\delta\ast}(\theta_i) \mathcal{M}_{\gamma\delta,\alpha\beta}(p_a,p_b;\theta_i) \phi_{A}^{\alpha}(\mathbf{p}_a) \phi_{B}^{\beta}(\mathbf{p}_b)\,,
\ee
and 
\be
Z(g^2) g^2\,\tilde{\phi}_A^{\gamma\ast}(\mathbf{p}_i) \tilde{\phi}_B^{\delta\ast}(\mathbf{p}_j) =\sum_{\epsilon\sigma} \int d(\cos\theta_c)\,\mathcal{M}^{\ast}_{\gamma\delta,\epsilon\sigma}(\theta_c; p_i,p_j)  \phi_{A}^{\epsilon\ast}(\theta_c) \phi_{B}^{\sigma\ast}(\theta_c)\,.
\ee

In the case of maximal wave-packet overlap 
one can equate, order by order, \refeq{eq:ot_wave_mod} to \refeq{eq:amplsqua}  
and compute $\Delta_{\textrm{max}}$
after applying the optical theorem. One finds, at the lowest order, 
\be\label{eq:max_over}
2\,\textrm{Im}\, Z(g^2) g^2 =\Delta_{\textrm{max}}\, g^4 = \frac{1}{16\pi} \int d(\cos\theta_i) \left|\mathcal{M}(p_A,p_B;\theta_i)\right|^2\,.
\ee
Changing the solid-angle distribution of the amplitude $\mathcal{M}(\theta_i)$ thus results in different values of $\Delta_{\textrm{max}}$.

\section{Eigenvalues of the reduced density matrix}\label{app:2eig}

In this appendix, we are going to prove that the density matrix, $\rho_Q$, has, in a perturbative setting, at most three non-zero eigenvalues, $e_1$,  $e_2$, $e_3$, which satisfy the condition $e_1\ll e_2, e_3$. 

The first step is to verify that $\textrm{det}(\rho_Q)=0$. 
To calculate the determinant of $\rho_Q$, we employ a well-known consequence of the Cayley-Hamilton theorem, which allows one to express the determinant of a matrix as a polynomial of its traces. For a $4\times 4$ matrix $A$ it reads:
\be\label{eq:det}
\textrm{det}(A)=\frac{1}{24}\Big([\textrm{Tr}(A)]^4-6\,\textrm{Tr}(A^2)[\textrm{Tr}(A)]^2+3\,[\textrm{Tr}(A^2)]^2+8\,\textrm{Tr}(A^3)\textrm{Tr}(A)-6\,\textrm{Tr}(A^4)\Big)\,.
\ee
Given that $\textrm{Tr}(\rho_Q)=1$, \refeq{eq:det} reduces to
\be\label{eq:detq}
\textrm{det}(\rho_Q)=\frac{1}{24}\Big(1-6\,\textrm{Tr}(\rho_Q^2)+3\,[\textrm{Tr}(\rho_Q^2)]^2+8\,\textrm{Tr}(\rho_Q^3)-6\textrm{Tr}(\rho_Q^4)\Big)\,.
\ee

Let us introduce a short-hand notation for \refeq{rho:flav}:
\be\label{eq:shorth}
\rho_Q=\pa+\Delta\left(i R_1+R_2 \right).
\ee
At the order $\mathcal{O}(\mathcal{M}^2)$ one finds
\bea
\textrm{Tr}(\rho_Q^2)&=&1+2i\Delta\textrm{Tr}(\pa R_1)+2\Delta\textrm{Tr}(\pa R_2)-\Delta^2\textrm{Tr}(R_1^2)\,,\label{eq:eq1} \\
\textrm{Tr}(\rho_Q^3)&=&1+3i\Delta\textrm{Tr}(\pa R_1)+3\Delta\textrm{Tr}(\pa R_2)-3\Delta^2\textrm{Tr}(\pa R_1^2)\,, \label{eq:eq2}  \\
\textrm{Tr}(\rho_Q^4)&=&1+4i\Delta\textrm{Tr}(\pa R_1)+4\Delta\textrm{Tr}(\pa R_2)-4\Delta^2\textrm{Tr}(\pa R_1^2)\nonumber\\
 & &-\,2\Delta^2[\textrm{Tr}(\pa R_1)]^2.\label{eq:eq3}
\eea
Plugging Eqs.~(\ref{eq:eq1})-(\ref{eq:eq3}) into \refeq{eq:detq} confirms that $\textrm{det}(\rho_Q)=0$.

Let us now show that $e_1\ll e_2, e_3$. 
To this end, let us introduce the auxiliary variables
\be
q_1=e_1e_2+e_1e_3+e_2e_3\,,\qquad q_2=e_1e_2e_3\,,
\ee
and verify that $q_2/q_1\ll1$. 
Using the traces
\be\label{eq:trq2q3}
\textrm{Tr}(\rho_Q^2)= e_1^2+e_2^2+e_3^2\,,\qquad \textrm{Tr}(\rho_Q^3)= e_1^3+e_2^3 +e_3^3\,,
\ee
one gets
\be\label{eq:qs}
q_1=\frac{1}{2}\left[1-\textrm{Tr}(\rho_Q^2)\right],\qquad q_2=\frac{1}{6}\left[1-3\,\textrm{Tr}(\rho_Q^2)+2\,\textrm{Tr}(\rho_Q^3)\right].
\ee

Substituting Eqs.~(\ref{eq:eq1}) and (\ref{eq:eq2}) 
into \refeq{eq:qs} and extracting $R_1$ explicitly from \refeq{eq:shorth} and \refeq{rho:flav}, one finds 
\be\label{eq:q2}
q_2=-\frac{\Delta^2}{2}\left(\abra\mfw\aket-\abra\mfw^\dagger\aket\right)^2 = 2\,\Delta^2 \abra \textrm{Im}\,\mfw \aket^2\,.
\ee
Since, by \refeq{eq:qs},
\be
q_1=\frac{1}{2}\,\mathcal{E}\propto 2\,\Delta \abra \textrm{Im}\,\mfw \aket\,,
\ee
$q_2$ in \refeq{eq:q2} is perturbatively suppressed with respect to $q_1$, which implies $q_2/q_1\ll 1$.

There is one special situation when $\Delta=\Delta_{\textrm{max}}$. This corresponds to a pure 
$\rho_Q$ matrix, $\textrm{Tr}(\rho_Q^2)=1$, which means that $\rho_Q$ has only one non-zero eigenvalue. 

\section{Derivation of the concurrence}\label{app:con}

In this appendix, we derive an analytic expression for the concurrence of 
the momentum-reduced density matrix $\rho_Q$ in terms of the scattering amplitude~$\mathcal{M}$ and the initial state~\aket. 

We start by recalling \refeq{eq:contr}:
\be\label{eq:contr_app}
C^2(\rho_Q)=\textrm{Tr}(R_Q)-\sqrt{2[\textrm{Tr}(R_Q)^2-\textrm{Tr}(R_Q^2)]}\,.
\ee
To facilitate the calculations, let us introduce the short-hand notation
\be\label{eq:short}
\rho_Q=\pa+\Delta\left(i R_1+R_2 \right), \qquad \tilde{\rho}_Q=\mathcal{P}_B+\Delta\left(i S_1+S_2 \right).
\ee
Using \refeq{eq:short}, the $R_Q$ matrix can be rewritten as
\begin{multline}
R_Q=\pa\prb + i\Delta\left(\pa S_1+R_1\prb\right)+\Delta\left(\pa S_2+R_2\prb\right) \\
+ i\Delta^2\left(R_1S_2+R_2S_1\right)+\Delta^2\left(R_2S_2
-R_1S_1\right).
\end{multline}
We thus have
\begin{multline}
\textrm{Tr}(R_Q)=|c_A|^2+ i\Delta\left(\abra S_1\aket+ \bbra R_1\bket\right)+\Delta\left(\abra S_2\aket+\bbra R_2\bket\right) \nonumber\\
+ i\Delta^2\,\textrm{Tr}\left(R_1S_2+R_2S_1\right)+\Delta^2\,\textrm{Tr}\left(R_2S_2
-R_1S_1\right),
\end{multline}
where $c_A$ is defined in \refeq{eq:cadef}. 

Let us next compute $\textrm{Tr}(R_Q)^2$ and $\textrm{Tr}(R_Q^2)$ at the order $\mathcal{O}(\mathcal{M}^4)$\,:
\bea\label{eq:trR2}
\textrm{Tr}(R_Q)^2&=&|c_A|^4+2i\Delta\,|c_A|^2\left(\abra S_1\aket+ \bbra R_1\bket\right)+2\Delta\,|c_A|^2\left(\abra S_2\aket+ \bbra R_2\bket\right)\nonumber\\
 & &+\,\Delta^2\left[\abra S_2\aket^2+\bbra R_2\bket^2-\abra S_1\aket^2-\bbra R_1\bket^2\right.\nonumber\\
 & &+\left. 2\abra S_2\aket\bbra R_2\bket-2\abra S_1\aket\bbra R_1\bket+2\,|c_A|^2\textrm{Tr}(R_2S_2-R_1S_1)\right]\nonumber\\
 & &+\,2i\Delta^2\left[\left(\abra S_1\aket+\bbra R_1\bket\right)\left(\abra S_2\aket+\bbra R_2\bket\right)+|c_A|^2\textrm{Tr}(R_1S_2+R_2S_1)\right]\nonumber\\
 & &-\,2\Delta^3\left[\left(\abra S_1\aket+\bbra R_1\bket\right)\textrm{Tr}(R_1S_2+R_2S_1)+\left(\abra S_2\aket+\bbra R_2\bket\right)\textrm{Tr}(R_1S_1)\right]\nonumber\\
 & &-\,2i\Delta^3\left(\abra S_1\aket+\bbra R_1\bket\right)\textrm{Tr}(R_1S_1)\nonumber\\
 & &+\,\Delta^4\,\textrm{Tr}(R_1S_1)^2
\eea
and
\bea\label{eq:tr2R}
\textrm{Tr}(R_Q^2)&=&|c_A|^2+2i\Delta\,|c_A|^2\big(\abra S_1\aket+ \bbra R_1\bket\big)+2\Delta\,|c_A|^2\left(\abra S_2\aket+ \bbra R_2\bket\right)\nonumber\\
 & &+\,\Delta^2\left[\textrm{Tr}(\pa S_2 \pa S_2)+\textrm{Tr}(\prb R_2\prb R_2)-\textrm{Tr}(\pa S_1\pa S_1)-\textrm{Tr}(\prb R_1\prb R_1)\right.\nonumber\\
 & &+\left. 2\,\textrm{Tr}(\pa S_2R_2 \prb)-2\,\textrm{Tr}(\pa S_1R_1\prb)+2\,\textrm{Tr}(\pa \prb R_2S_2)-2\,\textrm{Tr}(\pa \prb R_1S_1)\right]\nonumber\\
 & &+\,2i\Delta^2\left[\textrm{Tr}(\pa \prb R_1S_2)+\textrm{Tr}(\pa\prb R_2S_1)+\textrm{Tr}(\pa S_1\pa S_2)\right.\nonumber\\
 & &+\left.\textrm{Tr}(\pa S_1R_2\prb )+\textrm{Tr}(R_1\prb \pa S_2)+\textrm{Tr}(R_1\prb R_2\prb)\right]\nonumber\\
 & &-\,2\Delta^3\left[\textrm{Tr}(\pa S_1+R_1\prb)\textrm{Tr}(R_1S_2+R_2S_1)+\textrm{Tr}(\pa S_2+R_2\prb)\textrm{Tr}(R_1S_1)\right]\nonumber\\
 & &-\,2i\Delta^3\left[\textrm{Tr}(\pa S_1R_1S_1)+\textrm{Tr}(\prb R_1S_1R_1)\right]\nonumber\\
 & &+\,\Delta^4\,\textrm{Tr}(S_1R_1S_1R_1)\,.
\eea

The calculation of the traces is tedious yet straightforward. One obtains
\begin{align}
\abra S_1\aket=\bbra R_1\bket={}&-c_A d_1+c_A^\ast d_1^\ast\\
\abra S_2\aket=\bbra R_2\bket={}&f_1\\
\textrm{Tr}(R_1S_1)={}&-2|d_1|^2+c_Ad_3+c_A^\ast d_3^\ast \\
\textrm{Tr}(R_2S_2)={}&h\\   
\textrm{Tr}(R_1S_2)=\textrm{Tr}(R_2S_1)={}&f_2-f_2^\ast \\
\textrm{Tr}(R_0S_0R_1S_1)={}&-2c_A c^\ast_A|d_1|^2 + c^{\ast\,2}_A d_1^{\ast\,2} + c_A^2c^\ast_A d_3\\
\textrm{Tr}(S_0R_0S_1R_1)={}&-2c_A c^\ast_A|d_1|^2 + c^{2}_A d_1^{2} + c_A c^{\ast\,2}_A d_3^\ast\\
\textrm{Tr}(R_0S_1R_0S_1)=\textrm{Tr}(S_0R_1S_0R_1)={}&c_A^2 d_1^2 + c_A^{\ast\,2} d_1^2-2c_Ac_A^\ast|d_1|^2\\
\textrm{Tr}(R_0S_0R_2S_2)={}&c_A^\ast k\\
\textrm{Tr}(S_0R_0S_2R_2)={}&c_A k^\ast\\
\textrm{Tr}(R_0S_2R_0S_2)=\textrm{Tr}(S_0R_2S_0R_2)={}&f_1^2\\
\textrm{Tr}(R_0S_0R_1S_2)=\textrm{Tr}(R_0S_0R_2S_1)={}&c_A^\ast d_1^\ast f_1-c_Ac^\ast_Af_2^\ast\\
\textrm{Tr}(S_0R_0S_1R_2)=\textrm{Tr}(S_0R_0S_2R_1)={}&-c_A d_1 f_1+c_Ac^\ast_Af_2\\
\textrm{Tr}(R_0S_1R_0S_2)=\textrm{Tr}(S_0R_1S_0R_2)={}&c_A^\ast d_1^\ast f_1-c_A d_1 f_1\\
\begin{split}
\textrm{Tr}(R_0S_1R_1S_1)=-\textrm{Tr}(S_0R_1S_1R_1)={}&2c_A|d_1|^2d_1 + 2c_A^\ast |d_1|^2 d_1^\ast + c_A^2 d_1 d_3+ c_A c^\ast_A d_1 d_3^\ast\\
&+c_A^{\ast\,2} d_3^\ast d_1^\ast+c_A c^\ast_A d_1^\ast d_3
\end{split}\\
\textrm{Tr}(R_0S_1R_1S_2)=\textrm{Tr}(S_0R_2S_1R_1)={}&-|d_1|^2f_1 + c_A f_1d_3^\ast + c_A d_1f_2^\ast-c_A ^\ast d_1^\ast f_2^\ast\\
\textrm{Tr}(R_0S_1R_2S_1)=\textrm{Tr}(S_0R_1S_2R_1)={}&-|d_1|^2f_1 + c_A d_1f_2^\ast+c_A^\ast d_1^\ast f_2 -c_A c^\ast_A g\\
\textrm{Tr}(R_0S_2R_1S_1)={}&-|d_1|^2f_1+c_A f_1d_3+c_A^\ast d_1^\ast f_2^\ast-c_A d_1f_2^\ast\\
\textrm{Tr}(S_0R_1S_1R_2)={}&-|d_1|^2f_1+c_A f_1d_3+c_A^\ast d_1^\ast f_2-c_A d_1f_2\\
\begin{split}
\textrm{Tr}(R_1S_1R_1S_1)={}&2|d_1|^4 + c_A^{\ast\,2} d_3^{\ast\,2} + c_A^2 d_3^2-4 c_A d_3|d_1|^2-4 c_A^\ast d_3^\ast |d_1|^2\\
&+2c_A d_1^2d_3^\ast +2c_A^\ast d_1^{\ast\,2}d_3\,,
\end{split}
\end{align}
where we have defined
\begin{align}
d_1={}&\abra\mfw^\dagger\bket\\
d_2={}&\int d\Pi_2\,\bbra\mathcal{M}^{\dagger}\mathcal{M}\aket\\
d_3={}&\abra\mfw^\dagger\mtilfw\bket\\
f_1={}&\int d\Pi_2\,\bbra\mathcal{M}\pa\mathcal{M}^\dagger\bket\\
f_2={}&\bbra\mtilfw^\dagger\int d\Pi_2(\mathcal{M}\pa\mathcal{M}^\dagger)\bket\\
g={}&\bbra\mtilfw^\dagger\int d\Pi_2(\mathcal{M}\pa\mathcal{M}^\dagger)\mtilfw\bket\\
h={}&\textrm{Tr}\big[\int d\Pi_2(\mathcal{M}\pa\mathcal{M}^\dagger)\int d\Pi_2(\mtil\prb\mtil^\dagger)\big]\\
k={}&\bbra\int d\Pi_2(\mathcal{M}\pa\mathcal{M}^\dagger)\int d\Pi_2(\mtil\prb\mtil^\dagger)\aket\,.
\end{align}

It is now easy to verify that the terms of order 
$\mathcal{O}(1)$, $\mathcal{O}(\Delta)$, $\mathcal{O}(i\Delta^2)$, $\mathcal{O}(i\Delta^3)$ cancel out between \refeq{eq:trR2} and \refeq{eq:tr2R}. The term of order 
$\mathcal{O}(i\Delta)$ can be recast, using the optical theorem, as
\be
i\left(c_A^\ast d_1^\ast-c_A d_1\right)=-c_A d_2\,.
\ee

This finally allows us to rewrite \refeq{eq:contr_app} in a relatively compact form,
\begin{multline}\label{eq:consym_app}
C^2(\rho_Q)=|c_A|^2+2\Delta\left(f_1- c_A d_2\right)
+\Delta^2(2|d_1|^2-c_A d_3-c_A^\ast d_3^\ast)\\
-2\Delta\Big[f_1^2+|c_A|^2 h - (c_A^\ast k + c_A k^\ast)+\Delta^2|d_1^2 - c_A^\ast d_3|^2\\
-2\Delta\left(|d_1|^2f_1+c_A c^\ast_A g-c_A d_1f_2-c_A^\ast d_1^\ast f_2^\ast\right)\Big]^{1/2}.
\end{multline}

\bibliographystyle{JHEP}
\bibliography{mybib}

\end{document}

%% file: macros.tex
\newcommand{\newc}{\newcommand*}

\long\def\begincomment#1\endcomment{%
        \begingroup\sf\baselineskip12pt#1\endgroup}

\newc{\etal}{\textrm{et al.}} 
\newc{\eg}{\textrm{e.g.}} 
\newc{\ie}{\textrm{i.e.}}
\newc{\etc}{\textrm{etc}}
\newc\vs{\textrm{vs.}}
\newc{\cl}{\rm {C.L.}}

\newc{\ev}{\ensuremath{\,\mathrm{eV}}}
\newc{\kev}{\ensuremath{\,\mathrm{keV}}}
\newc{\mev}{\ensuremath{\,\mathrm{MeV}}}
\newc{\gev}{\ensuremath{\,\mathrm{GeV}}}
\newc{\tev}{\ensuremath{\,\mathrm{TeV}}}
\newc{\MeV}{\mev} 
\newc{\TeV}{\tev}
\newc{\invpb}{\ensuremath{/\text{pb}}}
\newc{\invfb}{\ensuremath{\,\textrm{fb}^{-1}}}
\newc\nb{\ensuremath{\,\mathrm{nb}}} \newc\pb{\ensuremath{\,\mathrm{pb}}} \newc\fb{\ensuremath{\,\mathrm{fb}}}
\newc\pc{\ensuremath{\,\mathrm{pc}}}
\newc\kpc{\ensuremath{\,\mathrm{kpc}}}
\newc\mpc{\ensuremath{\,\mathrm{Mpc}}}
\newc\ps{\ensuremath{\,\mathrm{ps}}} 
\newc\cmeter{\ensuremath{\,\mathrm{cm}}} 
\newc\meter{\ensuremath{\,\mathrm{m}}} 
\newc\kmeter{\ensuremath{\,\mathrm{km}}}
\newc\second{\ensuremath{\,\mathrm{s}}}
\newc\msecond{\ensuremath{\,\mathrm{ms}}}
\newc\nsecond{\ensuremath{\,\mathrm{ns}}}
\newc\psecond{\ensuremath{\,\mathrm{ps}}}

\newc{\chisqmin}{\ensuremath{\chi^2_{\mathrm{min}}}}
\newc{\Delchisq}{\ensuremath{\Delta\chi^2}}
\newc{\chisq}{\ensuremath{\chi^2}}
\newc{\like}{\ensuremath{\mathcal{L}}}

\newc\lsim{\ensuremath{\mathrel{\rlap{\lower4pt\hbox{\hskip1pt$\sim$}}\raise1pt\hbox{$<$}}}}
\newc\gsim{\ensuremath{\mathrel{\rlap{\lower4pt\hbox{\hskip1pt$\sim$}}\raise1pt\hbox{$>$}}}}
\newc{\VEV}[1]{\ensuremath{\langle #1 \rangle}}
\newc{\dl}{\ensuremath{\stackrel{\leftarrow}{D}}}
\newc{\dr}{\ensuremath{\stackrel{\rightarrow}{D}}}

\newc{\bcenter}{\begin{center}}    \newc{\ecenter}{\end{center}}
\newc{\bfl}{\begin{flushleft}}    \newc{\efl}{\end{flushleft}}
\newc{\bfr}{\begin{flushright}}    \newc{\efr}{\end{flushright}}

\newc{\bi}{\begin{itemize}}
\newc{\ei}{\end{itemize}}
\newc{\bed}{\begin{description}}
\newc{\eed}{\end{description}}
\newc{\ben}{\begin{enumerate}}
\newc{\een}{\end{enumerate}}

\newc{\be}{\begin{equation}}
\newc{\ee}{\end{equation}}
\newc{\bea}{\begin{eqnarray}}
\newc{\eea}{\end{eqnarray}}
\newc{\bfle}{\begin{flalign}}
\newc{\efle}{\end{flalign}}
\newc{\ra}{\rightarrow}

\newc{\alphas}{\ensuremath{\alpha_s}}
\newc{\alphatwo}{\ensuremath{\alpha_2}}
\newc{\alphaone}{\ensuremath{\alpha_1}}
\newc{\alphai}[1]{\ensuremath{\alpha_{#1}}}
\newc{\alphaem}{\ensuremath{\alpha_{\mathrm{em}}}}
\newc{\alphaeff}{\ensuremath{\alpha_{\mathrm{eff}}}}
\newc{\sineff}{\ensuremath{\sin \theta_{\mathrm{eff}}}}
\newc{\sinsqeff}{\ensuremath{\sin^2 \theta_{\mathrm{eff}}}}
\newc{\dalphahad}{\ensuremath{\Delta \alpha_{\mathrm{had}}}}
\newc{\yt}{\ensuremath{h_t}} \newc{\yb}{\ensuremath{h_b}} \newc{\ytau}{\ensuremath{h_{\tau}}}
\newc\mz{\ensuremath{M_Z}} 
\newc\mw{\ensuremath{m_W}}
\newc\mZ{\mz}        \newc\mW{\mw}
\newc\mhsm{\ensuremath{ m_{H_{\mathrm{SM}}}}}
\newc{\mtop}{\ensuremath{ m_t}}               \newc{\mtpole}{\ensuremath{ M_t}}
\newc{\mbottom}{\ensuremath{ m_b}} 
\newc{\mtau}{\ensuremath{ m_{\tau}}}
\newc{\mt}{\mtpole}
\newc{\mb}{\mbottom} 
\newc{\rtwogg}{\ensuremath{R_{h_2}(\gamma\gamma)}}
\newc{\rtwozz}{\ensuremath{R_{h_2}(ZZ)}}
\newc{\ronegg}{\ensuremath{R_{h_1}(\gamma\gamma)}}
\newc{\ronezz}{\ensuremath{R_{h_1}(ZZ)}}
\newc{\rsiggg}{\ensuremath{R_{h_\textrm{sig}}(\gamma\gamma)}}
\newc{\rsigzz}{\ensuremath{R_{h_\textrm{sig}}(ZZ)}}
\newc{\llbar}{\ensuremath{\ell\bar{\ell}}}
\newc{\tauptaum}{\ensuremath{ \tau^+\tau^-}}
\newc{\qqbar}{\ensuremath{ q\bar{q}}} \newc{\ppbar}{\ensuremath{ p\bar{p}}}
\newc{\bbbar}{\ensuremath{ b\bar{b}}} \newc{\ttbar}{\ensuremath{ t\bar{t}}}
\newc{\ffbar}{\ensuremath{ f\bar{f}}} \newc{\tautaubar}{\ensuremath{ \tau\bar{\tau}}}

\newc{\mchi}{\ensuremath{m_\neutone}}
\newc{\squark}{\ensuremath{\tilde{q}}}
\newc{\slepton}{\ensuremath{\tilde{l}}}
\newc{\gluino}{\ensuremath{\tilde{g}}} 
\newc{\mgluino}{\ensuremath{{m_{\gluino}}}}
\newc{\wino}{\ensuremath{\tilde{W}}} 
\newc{\mwino}{\ensuremath{{m_{\wino}}}}
\newc{\tone}{\ensuremath{{\tilde{t}_1}}}
\newc{\bone}{\ensuremath{{\tilde{b}_1}}}
\newc{\Hone}{\ensuremath{{\tilde{H}_{1}}}}
\newc{\Htwo}{\ensuremath{{\tilde{H}_{2}}}}
\newc{\Hhtwo}{\ensuremath{{H_{2}}}}
\newc{\qli}{\ensuremath{{\tilde{Q}_{i}}}}
\newc{\uri}{\ensuremath{{\tilde{u}_{i}}}}
\newc{\dri}{\ensuremath{{\tilde{d}_{i}}}}
\newc{\lli}{\ensuremath{{\tilde{L}_{i}}}}
\newc{\eri}{\ensuremath{{\tilde{e}_{i}}}}

\newc{\sthw}{\ensuremath{ \sin\theta_W}}              \newc{\cthw}{\ensuremath{\cos\theta_W}}
\newc{\tanthw}{\ensuremath{ \tan\theta_W}}              \newc{\cotthw}{\ensuremath{\cot\theta_W}}
\newc{\ssqthw}{\ensuremath{\sin^2 \theta_W}}
\newc{\msbar}{\ensuremath{\overline{MS}}} \newc{\drbar}{\ensuremath{\overline{DR}}}
\newc{\mtmtsmmsbar}{\ensuremath{ m_t(m_t)^{\msbar}_{{\mathrm{SM}}}}}
\newc{\mtmtsmdrbar}{\ensuremath{ m_t(m_t)^{\drbar}_{{\mathrm{SM}}}}}
\newc{\mtmtmssmdrbar}{\ensuremath{ m_t(m_t)^{\drbar}_{{\mathrm{SUSY}}}}}
\newc{\mbmbmsbar}{\ensuremath{ m_b(m_b)^{\msbar} }}
\newc{\mbmbsmmsbar}{\ensuremath{ m_b(m_b)^{\msbar}_{{\mathrm{SM}}}}}
\newc{\mbmzsmmsbar}{\ensuremath{ m_b(\mz)^{\msbar}_{{\mathrm{SM}}}}}
\newc{\mbmzsmdrbar}{\ensuremath{ m_b(\mz)^{\drbar}_{{\mathrm{SM}}}}}
\newc{\mbmzmssmdrbar}{\ensuremath{ m_b(\mz)^{\drbar}_{{\mathrm{SUSY}}}}}
\newc{\mtaumzsmmsbar}{\ensuremath{ m_{\tau}(\mz)^{\msbar}_{{\mathrm{SM}}}}}
\newc{\mtaumzsmdrbar}{\ensuremath{ m_{\tau}(\mz)^{\drbar}_{{\mathrm{SM}}}}}
\newc{\mtaumzmssmdrbar}{\ensuremath{ m_{\tau}(\mz)^{\drbar}_{{\mathrm{SUSY}}}}}
\newc{\alphasmzms}{\ensuremath{\alpha_s(M_Z)^{\overline{MS}}}}
\newc{\alphaimzms}[1]{\ensuremath{\alpha_{#1}(M_Z)^{\overline{MS}}}}
\newc{\alphaemmz}{\ensuremath{\alpha_{\mathrm{em}}(M_Z)^{\overline{MS}}}}

\newc{\mzero}{\ensuremath{{m_0}}}
\newc{\mhalf}{\ensuremath{ m_{1/2}}}
\newc{\tanb}{\ensuremath{\tan\beta}}
\newc{\azero}{\ensuremath{ A_0}}
\newc{\signmu}{\ensuremath{\rm{sgn}\,\mu}}
\newc{\atau}{\ensuremath{{A_{\tau}}}}
\newc{\mueff}{\ensuremath{\mu_{\rm{eff}}}}
\newc{\lam}{\ensuremath{{\lambda}}}
\newc{\kap}{\ensuremath{{\kappa}}}
\newc{\alam}{\ensuremath{{A_{\lambda}}}}
\newc{\akap}{\ensuremath{{A_{\kappa}}}}
\newc{\hs}{\ensuremath{ H_s}}      
\newc{\mhs}{\ensuremath{ m_{H_s}}} 
\newc{\mgut}{\ensuremath{ M_{\rm GUT}}}
\newc{\gut}{\ensuremath{{\rm GUT}}}
\newc{\mplanck}{\ensuremath{ M_{\rm P}}}      \newc{\mpl}{\ensuremath{ M_{\rm Pl}}}
\newc{\msusy}{\ensuremath{ M_{\rm SUSY}}}      \newc{\ms}{\ensuremath{ M_{\rm S}}}
 \newc{\hu}{\ensuremath{ H_u}}       \newc{\hd}{\ensuremath{ H_d}}
 \newc{\mhu}{\ensuremath{ m_{H_u}}}       \newc{\mhd}{\ensuremath{ m_{H_d}}}
 \newc{\mhuew}{\ensuremath{ m^{\ast}_{H_u}}}       \newc{\mhdew}{\ensuremath{ m^{\ast}_{H_d}}}
 \newc{\mhuewsq}{\ensuremath{ m^{\ast\, 2}_{H_u}}}       \newc{\mhdewsq}{\ensuremath{ m^{\ast\, 2}_{H_d}}}
 \newc{\mhl}{\ensuremath{m_\hl}} 
 \newc{\mhone}{\ensuremath{m_{h_1}}} 
 \newc{\mhtwo}{\ensuremath{m_{h_2}}} 
 \newc{\mhi}{\ensuremath{m_{\tilde{h}}}} 
 \newc{\mul}{\ensuremath{m_{\tilde{u}_L}}} 
 \newc{\mbone}{\ensuremath{m_{\tilde{b}_1}}}  
 \newc{\mtone}{\ensuremath{m_{\tilde{t}_1}}} 
 \newc{\ma}{\ensuremath{m_A}} 
 \newc{\mH}{\ensuremath{m_H}} 
 \newc{\maone}{\ensuremath{m_{a_1}}} 
 \newc{\matwo}{\ensuremath{m_{a_2}}}
 \newc{\hone}{\ensuremath{h_1}}
 \newc{\htwo}{\ensuremath{h_2}}
 \newc{\aone}{\ensuremath{a_1}}
 \newc{\atwo}{\ensuremath{a_2}}
 \newc{\mqthree}{\ensuremath{m_{\tilde{Q}_3}^2}}
 \newc{\muthree}{\ensuremath{m_{\tilde{u}_3}^2}}
 \newc{\mqli}{\ensuremath{m_{\tilde{Q}_{i}}}}
 \newc{\muri}{\ensuremath{m_{\tilde{u}_{i}}}}
 \newc{\mdri}{\ensuremath{m_{\tilde{d}_{i}}}}
 \newc{\mlli}{\ensuremath{m_{\tilde{L}_{i}}}}
 \newc{\meri}{\ensuremath{m_{\tilde{e}_{i}}}}
 \newc{\ts}{\ensuremath{T_{SUSY}}}

\newc{\sigsip}{\ensuremath{\sigma^{\rm SI}_{p}}}	\newc{\sigsin}{\ensuremath{\sigma^{\rm SI}_{n}}}
\newc{\sigsdp}{\ensuremath{\sigma^{\rm SD}_{p}}}	\newc{\sigsdn}{\ensuremath{\sigma^{\rm SD}_{n}}}
\newc{\sigsi}{\ensuremath{\sigma^{\rm SI}}}	\newc{\sigsd}{\ensuremath{\sigma^{\rm SD}}}
\newc{\abund}{\ensuremath{ \Omega h^2}}
\newc{\omegadm}{\ensuremath{ \Omega_{{\rm DM}}}}     \newc{\abunddm}{\ensuremath{ \Omega_{{\rm DM}} h^2}} 
\newc{\omegam}{\ensuremath{ \Omega_{{\rm m}}}}       \newc{\abundm}{\ensuremath{ \Omega_{{\rm m}} h^2}}
\newc{\omegab}{\ensuremath{ \Omega_{{\rm b}}}}	\newc{\abundb}{\ensuremath{ \Omega_{{\rm b}} h^2}}
\newc{\omegatot}{\ensuremath{ \Omega_{{\rm TOT}}}}
\newc{\omegacdm}{\ensuremath{ \Omega_{{\rm CDM}}}}   \newc{\abundcdm}{\ensuremath{ \Omega_{{\rm CDM}} h^2}}
\newc{\omegalambda}{\ensuremath{ \Omega_{\Lambda}}} \newc{\abundlambda}{\ensuremath{ \Omega_{\Lambda} h^2}}
\newc{\omegarad}{\ensuremath{ \Omega_{{\rm rad}}}}  \newc{\abundrad}{\ensuremath{ \Omega_{{\rm rad}} h^2}}
\newc{\rhocrit}{\ensuremath{ \rho_{\rm crit}}}
\newc{\rhochi}{\ensuremath{ \rho_{\chi}}}
\newc{\abunchi}{\ensuremath{\Omega_\chi h^2}}
\newc{\abundlsp}{\ensuremath{\Omega_{\rm LSP}h^2}}

\newc{\amu}{\ensuremath{ a_{\mu}}}        \newc{\amususy}{\ensuremath{ a_{\mu}^{\mathrm{SUSY}}}}
\newc{\amuexpt}{\ensuremath{ a_{\mu}^{\mathrm{expt}}}}        \newc{\amusm}{\ensuremath{ a_{\mu}^{\mathrm{SM}}}}
\newc\deltaamu{\ensuremath{\Delta a_{\mu}}} \newc{\deltaamususy}{\ensuremath{\delta a_{\mu}^{\mathrm{SUSY}}}}
\newc\gmtwo{\ensuremath{ (g-2)_{\mu}}} 
\newc{\deltagmtwomususy}{\ensuremath{\delta\left(g-2\right)_{\mu}^{\mathrm{SUSY}}}}
\newc{\deltagmtwomu}{\ensuremath{\delta\left(g-2\right)_{\mu}}}
\newc\BR{\ensuremath{\rm BR}}
\newc\bsgamma{\ensuremath{ b\rightarrow s \gamma }}
\newc\bxsgamma{\ensuremath{\overline{B}\rightarrow X_{s}\gamma}}
\newc\brbsgamma{\ensuremath{\BR\left(\bsgamma\right)}}
\newc\brbxsgamma{\ensuremath{\BR\left(\bxsgamma\right)}}
\newc\bsmumu{\ensuremath{B_s\to\mu^+\mu^-}}
\newc\brbsmumu{\ensuremath{\BR\left(B_s\to\mu^+\mu^-\right)}}
\newc\bdmmumu{\ensuremath{\overline{B}_d\to\mu^+\mu^-}}
\newc\bbbarmix{\ensuremath{\overline{B}_s\mbox{-}B_s}}      
\newc\delmbs{\ensuremath{\Delta M_{B_s}}}
\newc{\butaunu}{\ensuremath{B_u \rightarrow \tau \nu}}
\newc{\brbutaunu}{\ensuremath{\BR\left(B_u \rightarrow \tau \nu\right)}}

        \newcommand*{\refeq}[1]{Eq.~(\ref{#1})}
     \newcommand*{\refsec}[1]{Sec.~\ref{#1}}

\newcommand*{\neutone}{\ensuremath{\chi^0_1}}

\let\oldcite\cite
\renewcommand*{\cite}{~\oldcite}

\newcommand*{\hl}{\ensuremath{h}}

\newcommand*{\aket}{\ensuremath{|A\rangle}}
\newcommand*{\abra}{\ensuremath{\langle A|}}
\newcommand*{\bket}{\ensuremath{|B\rangle}}
\newcommand*{\bbra}{\ensuremath{\langle B|}}
\newcommand*{\pa}{\ensuremath{\mathcal{P}_{A}}}
\newcommand*{\prb}{\ensuremath{\mathcal{P}_{B}}}
\newcommand*{\mfw}{\ensuremath{\mathcal{M}_{\textrm{fw}}}}
\newcommand*{\mtil}{\ensuremath{\widetilde{\mathcal{M}}}}
\newcommand*{\mtilfw}{\ensuremath{\widetilde{\mathcal{M}}_{\textrm{fw}}}}
